\documentclass[preprint2]{emulateapj}

\newcommand{\kms}{$\,$km$\,$s$^{-1}$}

\usepackage{courier,graphicx,epsfig,color,colordvi,natbib,ifpdf}

\def\ud{UD}
\def\uds{UDs}

\def\Uds{UDs}
\def\vhigh{$v_{\rm high}$}
\def\vdeep{$v_{\rm deep}$}

\newcommand{\hirokoemail}{watanabe@kwasan.kyoto-u.ac.jp}

\begin{document}

\title{Temporal Evolution of  Velocity and Magnetic Field in and around Umbral Dots}

\author{Hiroko Watanabe}
\affil{Unit of Synergetic Studies for Space, %
      Kyoto University,  17 Kitakazan Ohmine-cho, 
      Yamashina-ku, %
      Kyoto 607-8417, JAPAN}
\affil{Kwasan and Hida Observatories, %
      Kyoto University,  Yamashina-ku, %
      Kyoto 607-8417, JAPAN; \hirokoemail}

\author{Luis R.~Bellot Rubio}
\affil{Instituto de Astrof\'{\i}sica de Andaluc\'{\i}a (CSIC), %
	Apartado de Correos 3004, 18080 Granada, SPAIN}

\author{Jaime de la Cruz Rodr\'iguez}
\affil{Department of Physics and Astronomy, Uppsala University, %
Box 516, SE-75120 Uppsala, SWEDEN}

\and

\author{Luc Rouppe van der Voort}
\affil{Institute of Theoretical Astrophysics,
  University of Oslo, %
  P.O. Box 1029 Blindern, N-0315 Oslo, NORWAY}

\shorttitle{Temporal Evolution of Umbral Dots}
\shortauthors{H.~Watanabe et al.}

\begin{abstract}
  We study the temporal evolution of umbral dots (\uds) using
  measurements from the CRISP imaging spectropolarimeter at the
  Swedish 1-m Solar Telescope.  Scans of the magnetically sensitive
  630~nm iron lines were performed under stable atmospheric conditions
  for 71~min with a cadence of 63~s.  These observations allow us to
  investigate the magnetic field and velocity in and around \uds\ at a
  resolution approaching 0\farcs13. From the analysis of 339 UDs, we
  draw the following conclusions: (1)~UDs show clear hints of upflows,
  as predicted by magnetohydrodynamic (MHD) simulations.  By contrast,
  we could not find systematic downflow signals. Only in very deep
  layers we detect localized downflows around \uds, but they do not
  persist in time.  (2) We confirm that UDs exhibit weaker and more
  inclined fields than their surroundings, as reported
  previously. However, \uds\ that have strong fields above 2000~G or
  are in the decay phase show enhanced and more vertical fields.  (3)
  There are enhanced fields at the migration front of \uds\ detached
  from penumbral grains, as if their motion were impeded by the
  ambient field.  (4) Long-lived \uds\ travel longer distances with
  slower proper motions.  Our results appear to confirm some aspects
  of recent numerical simulations of magnetoconvection in the umbra
  (e.g., the existence of upflows in UDs), but not others (e.g., the
  systematic weakening of the magnetic field at the position of UDs.)

\end{abstract}


\section{Introduction}\label{sec:introduction}

Umbral dots (\uds) are transient brightenings observed in sunspot
umbrae and pores, with typical sizes of 300~km and lifetimes of
10~min \citep[e.g.,][]{1997A&A...328..682S,
  1997A&A...328..689S}. 
They cover only 3--10\%\ of the umbral area, but contribute 10--20\%\
of its brightness.  For this reason, \uds\ are believed to play a vital role in the
energy balance of sunspots \citep{1965ApJ...141..548D, 
  1993ApJ...415..832S, 2010SoPh..267....1M}. 
  
\uds\ exhibit systematic proper motions in mature sunspots: those
appearing in the central umbral region are static, while \uds\ in
peripheral regions move inward with an average velocity of 1.0\,\kms\/
\citep{2007PASJ...59S.585K, 2008A&A...492..233R}.  Some peripheral
\uds\ are the continuation of penumbral grains---bright elongated
structures at the head of penumbral filaments that move toward the
center of the sunspot with speeds of about 0.4\,\kms\
\citep{1999A&A...348..621S, 2006ApJ...646..593R}.  
When the migration front detaches into a circular bright point, 
the tip of the penumbral grain becomes an \ud.

It is believed that the mechanism behind \uds\ is convection
interacting with the strong vertical field of the umbra, and many
observational results support this idea
\citep{2008ApJ...678L.157R, 2010A&A...510A..12B, 2010SoPh..266....5W}. 
In the formation phase of sunspots, \uds\ are akin to granules but
their apparent motion is more stochastic because of the surrounding
magnetic field \citep{1999ApJ...511..436S}. 
In developed sunspots, \uds\ are small and quiescent due to the 
stronger suppression of convection.

UD research is entering a new phase in which computer simulations
guide observational efforts.  The innovative simulations by
\citet{2006ApJ...641L..73S} predicted UDs with central dark lanes and
small localized downflow patches at their ends. A clear detection of
those features would immediately validate the numerical models, so
they have been the target of recent observations. The dark lanes are
the result of enhanced density in the upper central part of UDs,
caused by the piling up of hot gas that rises from deeper down. Once
the gas reaches the surface, it cools by radiative losses and descends
in narrow downflow channels at the end of the dark lanes.
\citet{2007ApJ...669L..57B} observed a dark lane in a big
\ud. However, the very large size of this UD ($>$1000\,km) suggests
that it could actually have been a cluster of several \uds. The first
detection of downflows surrounding an UD was presented by
\citet{2007ApJ...665L..79B}. Subsequently, \citet{2010ApJ...713.1282O} 
reported solid evidence of dark lanes and localized downflows based on
spectropolarimetric observations taken at the Swedish 1-m Solar
Telescope.  The sizes of the dark lanes and downflowing patches found
by \citet{2010ApJ...713.1282O} are near the diffraction limit of the
telescope, with the substructures keeping their identity for periods
of only a few minutes.  These authors also reported enhanced net
circular polarization at the site of the downflows.

\begin{figure*}[bhtp]
\centerline{\includegraphics[width=1\textwidth, bb=0 0 1020 651]{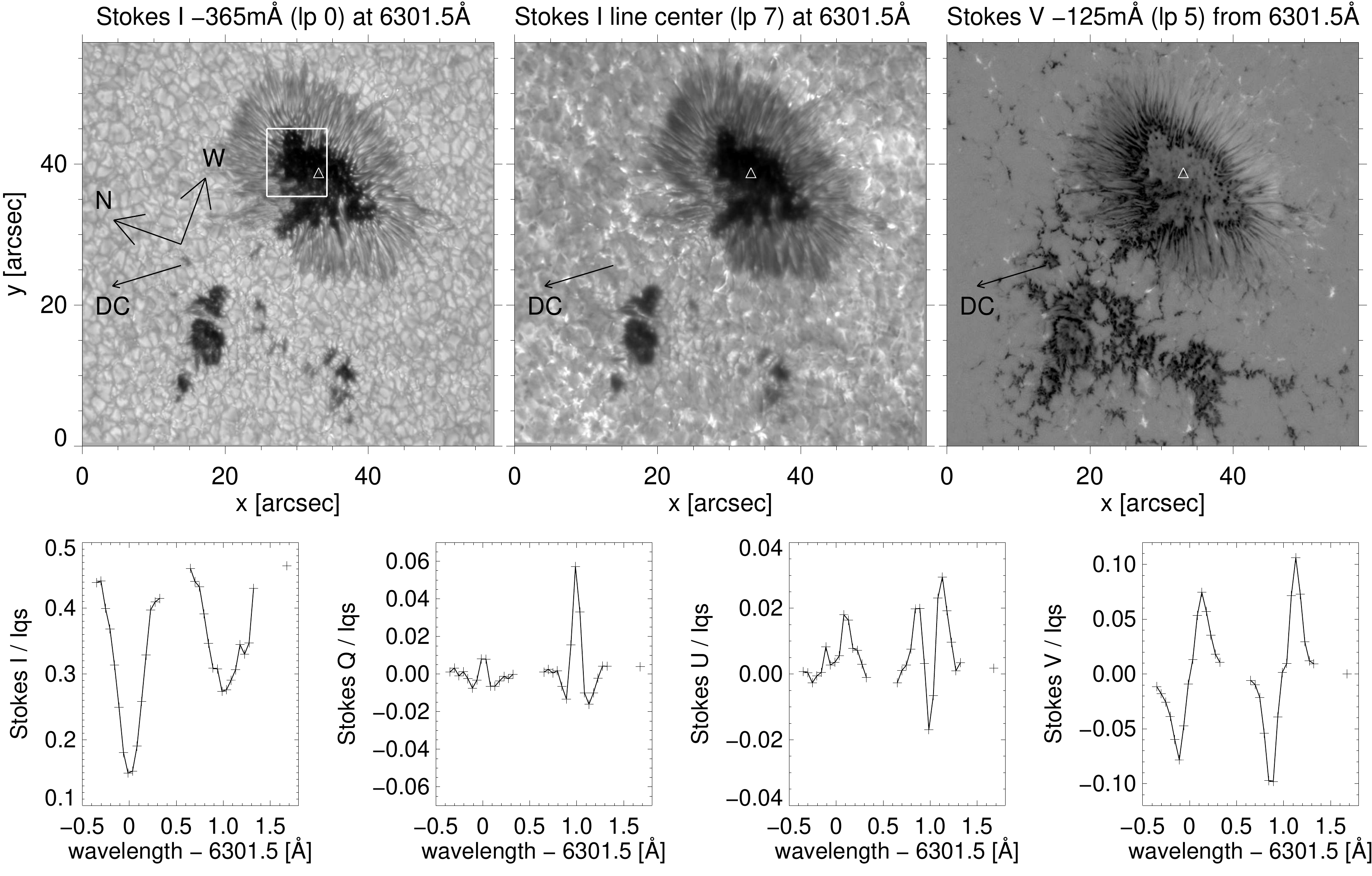}}
\caption{Filtergrams from the best scan of the data set, taken at
  08:30\,UT.  The full FOV is shown. From left to right: Stokes $I$ in
  the blue wing of 6301.5\,\AA\ (line position 0), Stokes $I$ at line
  center (line position 7), and Stokes $V$ in the blue wing
  (line position 5).  The direction to disk center (DC) is displayed
  with an arrow.  The white rectangle indicates the FOV of
  Figure~\ref{fig:cut_image}.  The bottom row shows the Stokes
  profiles emerging from the \ud\ marked with triangles in the upper
  panels. The $+$-symbols indicate the measured signals. }
    \label{fig:fig1}
\end{figure*}

The evolution of \uds\ and their magnetic fields is difficult to
study---and hence poorly known---because one needs full vector
spectropolarimetric measurements at very high temporal and spatial
resolution.  To the best of our knowledge, the magnetic properties of
\Uds\ have never been investigated at the required cadence and spatial
resolution \citep[but see][]{2009ApJ...694.1080S}.
\citet{2010ApJ...713.1282O} performed a preliminary analysis of 
the temporal evolution of six \uds, and this work should be considered
a substantial extension of their study.  Both the cadence and the
polarimetric sensitivity of our measurements are improved with respect
to those of \citet{2010ApJ...713.1282O}, as is the total duration of
the observations, 71 minutes, during which the seeing conditions were
excellent and stable. We use this unique data set to investigate the
evolution of the magnetic and velocity fields in and around \uds.

The paper is organized as follows.  The observations are described in
Section~\ref{sec:observation}, followed by an account of the methods
used for the detection of \uds\ and derivation of the velocity and
magnetic information (Section~\ref{sec:reduction}).  In
Section~\ref{sec:convection} we quantify how convection is modified in
the umbra.  In Section~\ref{sec:UDs} we describe the evolution of
some typical \uds, and present the results of our statistical analysis.
Finally, based on these results, we discuss the physical properties of
\uds\ in Section~\ref{sec:discussion}.


\section{Observations and Data Reduction}\label{sec:observation}

The observations were obtained with the CRisp Imaging
Spectro-Polarimeter (CRISP) at the Swedish 1-m Solar Telescope
\citep[SST;][]{2003SPIE.4853..341S}
on La Palma, Spain.  CRISP is based on a dual Fabry-P\'erot
interferometer similar to that described by
\citet{2006A&A...447.1111S}. The incoming light is modulated by two
liquid crystal variable retarders cycling through four states and then
analyzed by a polarizing beam splitter in front of two narrow-band
cameras. The narrow-band cameras record orthogonal polarization states
to minimize seeing-induced crosstalk.  CRISP has a third camera for
wide band imaging.  All the cameras operate at 35 frames per second and
take exposures of 17.6 ms. The synchronism between them is ensured by
an external optical chopper.

CRISP was used to measure the four Stokes profiles of the magnetically
sensitive \ion{Fe}{1}~6301.5 and 6302.5\,\AA\ lines, each sampled at
15 wavelength positions in steps of 48\,m\AA, from $-$350 to
$+$322\,m\AA.  Line positions 0--14 sample the 6301.5\,\AA\ line,
while positions 15--29 correspond to the 6302.5\,\AA\ line.  In
addition, a continuum wavelength point (6303.2\,\AA, line position 30)
was measured.  We recorded 9 frames per modulation state, resulting in
36 exposures per wavelength position.  The total time for a complete
wavelength scan of the two \ion{Fe}{1} lines plus the continuum point
was 32\,s.  Another 30\,s were needed to scan the \ion{Ca}{2} line at
8542~\AA\/ (not used in this paper).  Thus, the temporal cadence of the
\ion{Fe}{1} scans is 63\,s. The average CRISP transmission
profile has a Gaussian core (FWHM of 64 m\AA\/ at 6300\AA) and wide
Lorentzian wings. This transmission profile reduces the line core
depth of the \ion{Fe}{1} lines by about 20\%.

\begin{figure*}[t]
\centerline{\includegraphics[width=0.95\textwidth, bb=0 0 510 538]{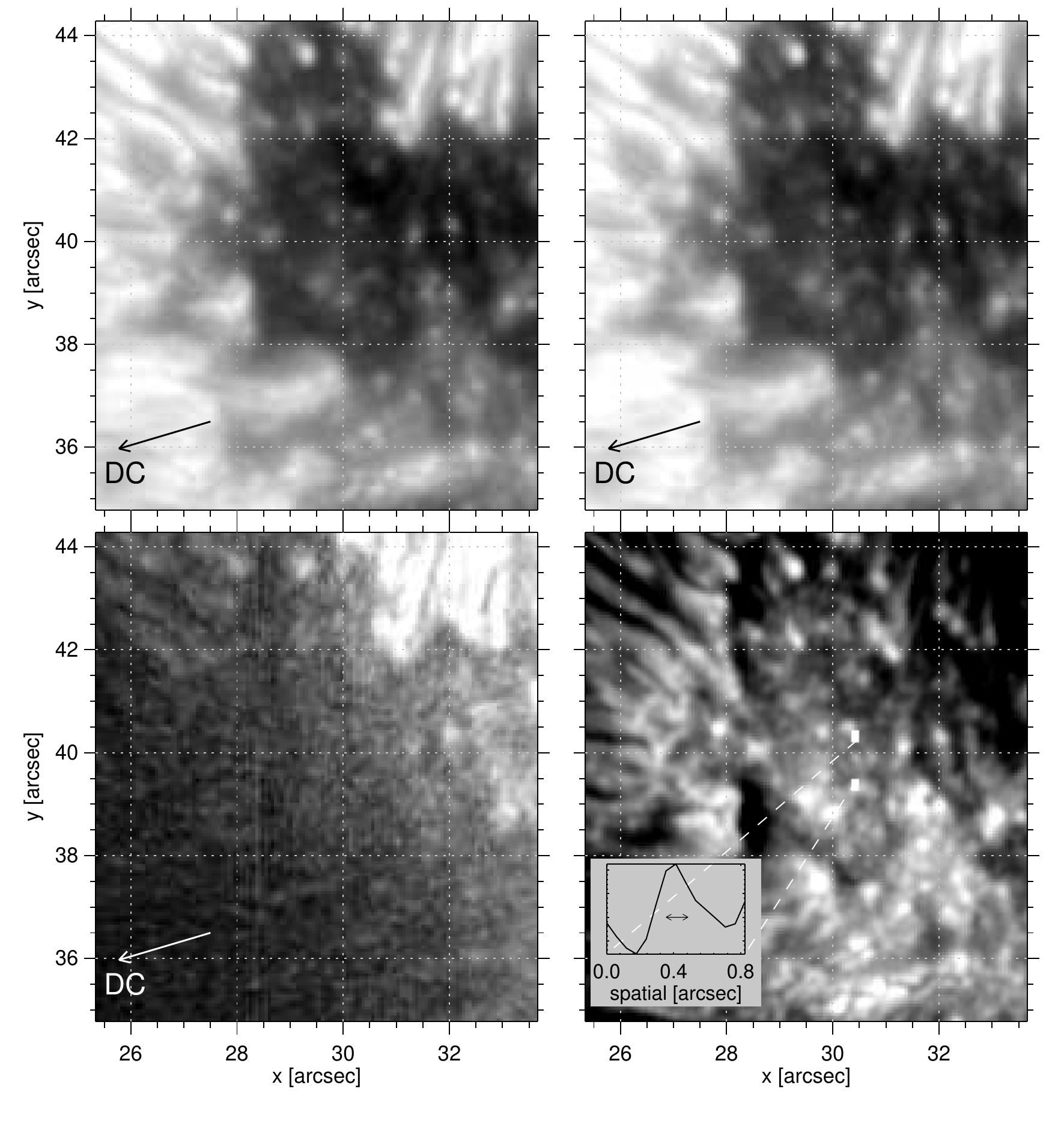}}
\vspace*{-1em}
\caption{Close-up of the region indicated with a white box in
  Figure~\ref{fig:fig1}. Clockwise, starting from upper left:
  intensity in the blue wing of \ion{Fe}{1} 6301.5\,\AA\ (line
  position 0), continuum intensity at 6303.2\,\AA\ (line position 30),
  CP map, and LP map.  The direction to disk center (DC) is displayed
  with an arrow.  A cut of a point-like feature (corresponding to an
  \ud) is shown in the CP map. The length of the arrow in the plot is
  0\farcs13 (equal to the diffraction limit).}  \label{fig:cut_image}
\end{figure*}

The CRISP etalons are mounted in tandem on a telecentric beam. The
separation of the cavities of the high resolution etalon sets the
wavelength of the transmission profile. This is not strictly the same
over the whole field of view (FOV) because the surface cannot be
infinitely flat, producing random wavelength shifts (cavity errors)
across the FOV. Intensity fluctuations, introduced by cavity errors in
the presence of a spectral line, were removed from the flat-field
images in a similar way as described by \citet{2011schnerr}. Their
flat-fielding scheme computes a cavity-error-free averaged quiet-sun
profile, which is removed from the flats on a pixel-by-pixel
basis. This average is obtained by summing many hundreds of exposures
acquired by moving the telescope in circles around the disk center.
The difference between the \citet{2011schnerr} scheme and ours is that
we applied a polarization-free flat for all the four Stokes states
taken at one wavelength position. 

We also corrected the data for spectral intensity gradients introduced
by the CRISP prefilter. The prefilter correction can be decomposed in
two contributions, an average prefilter shape and a term that accounts
for pixel-to-pixel deviations:
\begin{equation}
P(\lambda,x,y) = P_{\rm g}(\lambda) + {\rm d}P(x, y).
\end{equation}
The second term of the equation is included in the flat-field
correction. To obtain the global prefilter shape ($P_{\rm g}$), our
estimate of the averaged quiet-sun profile is compared to the FTS
atlas convolved with the CRISP transmission profile. The prefilter 
is assumed to have a Lorentzian shape which is multiplied with a
polynomial term to account for asymmetries:
\begin{equation}
P_{\rm g}(\lambda) = \frac{1}{1+[2(\lambda - \lambda_0)/w]^{2N_{\rm cav}}} \,(1 + p_0 \lambda + p_1 \lambda^3).
\end{equation}
Here, $\lambda_0$ and $w$ are the central wavelength and FWHM of the
prefilter, respectively, $N_{\rm cav}$ is the number of cavities of
the filter ($N_{\rm cav} =2$ in this case), and $p_0$ and $p_1$ are
the coefficients of the polynomial. A least-squares-fitting scheme
\citep{2009markwardt} was used to compute the prefilter parameters,
minimizing the differences between the observed and modeled
curves. All Stokes parameters were then corrected by dividing each
spectrum with the fitted prefilter curve.

The theoretical diffraction limit of the telescope around 6300\,\AA\
is 0\farcs13, and the image pixel size is 0\farcs059.  To ensure very
high spatial resolution, we used the adaptive optics system of the SST
and the Multi-Object Multi-Frame Blind Deconvolution technique
\citep[MOMFBD;][]{2005SoPh..228..191V}. The MOMFBD algorithm considers
all frames taken in one scan (31 wavelength points $\times$ 4
modulation states $\times$ 9 repetitions $\times$ 3 cameras) to remove
image distortions from the individual filtergrams.

Under enhanced differential-seeing conditions, residual rubber sheet
distortions are present between the narrow-band images of the same
scan even after MOMFBD restoration. This effect appears when the size
of the patches used to divide the images for the MOMFBD processing
(here $128 \times 128$ pixels) is larger than the spatial scale of the
atmospheric distortion (or iso-planatic patch). To attenuate their
influence in our measurements, we employed an extra step in the
processing following an idea from V.\ Henriques (private
communication). The wide-band images were used twice in a MOMFBD
restoration in the following manner:
\begin{enumerate}
\item All the frames were combined to produce the reference 
anchor image.  
\item The frames were separated in sets associated with each 
wavelength position and modulation state, resulting in one restored
wide-band image per wavelength and modulation state.
\end{enumerate}

The second set of wide-band images was not used for wavefront sensing,
i.e., they did not contribute to the determination of the Point Spread
Function in the MOMFBD calculations. The state-dependent restored
wide-band images were compared with the anchor wide-band image to
remove the residual rubber sheet deformations in the individual
filtergrams.  This correction was applied prior to the demodulation of
the data to achieve almost perfect co-alignment between the 4
modulation states from which the Stokes parameters are derived at any
wavelength position. The method also achieves almost perfect
co-alignment between the different wavelength positions to ensure the
integrity of the Stokes profiles over the FOV.

After restoration, the images were demodulated and corrected for
instrumental polarization using the telescope model developed by
\cite{2010selbing}. For details, see \cite{2008A&A...489..429V}.
In addition, we corrected small residual crosstalks from Stokes $I$ to
$Q$, $U$, and $V$ by forcing the polarization to be zero in the far
line wings.  All the Stokes profiles were normalized to the average
quiet-sun continuum intensity ($I_{\rm qs}$) computed for each time
step.  The typical noise levels in Stokes $Q$, $U$, and $V$ are
$1.9\times10^{-3}$, $2.8\times10^{-3}$, and $1.9\times10^{-3}$ 
of the continuum intensity, respectively.

We followed the main sunspot of NOAA active region 11024 from 08:05 to
09:16\,UT on 6 July 2009. The spot was located at 25$^{\circ}$S and
23$^{\circ}$W (heliocentric angle of 36$^{\circ}$, $\mu$=0.81).
During the observations the atmospheric conditions were excellent and
stable.  Figure~\ref{fig:fig1} shows selected filtergrams from one of
the best scans (08:30\,UT) and the four Stokes profiles measured at
the position of an \ud. The size of the full FOV is $58\arcsec \times
57\arcsec$.

\begin{figure*}[t]
\centerline{\includegraphics[width=1\textwidth, bb=0 0 1020 977]{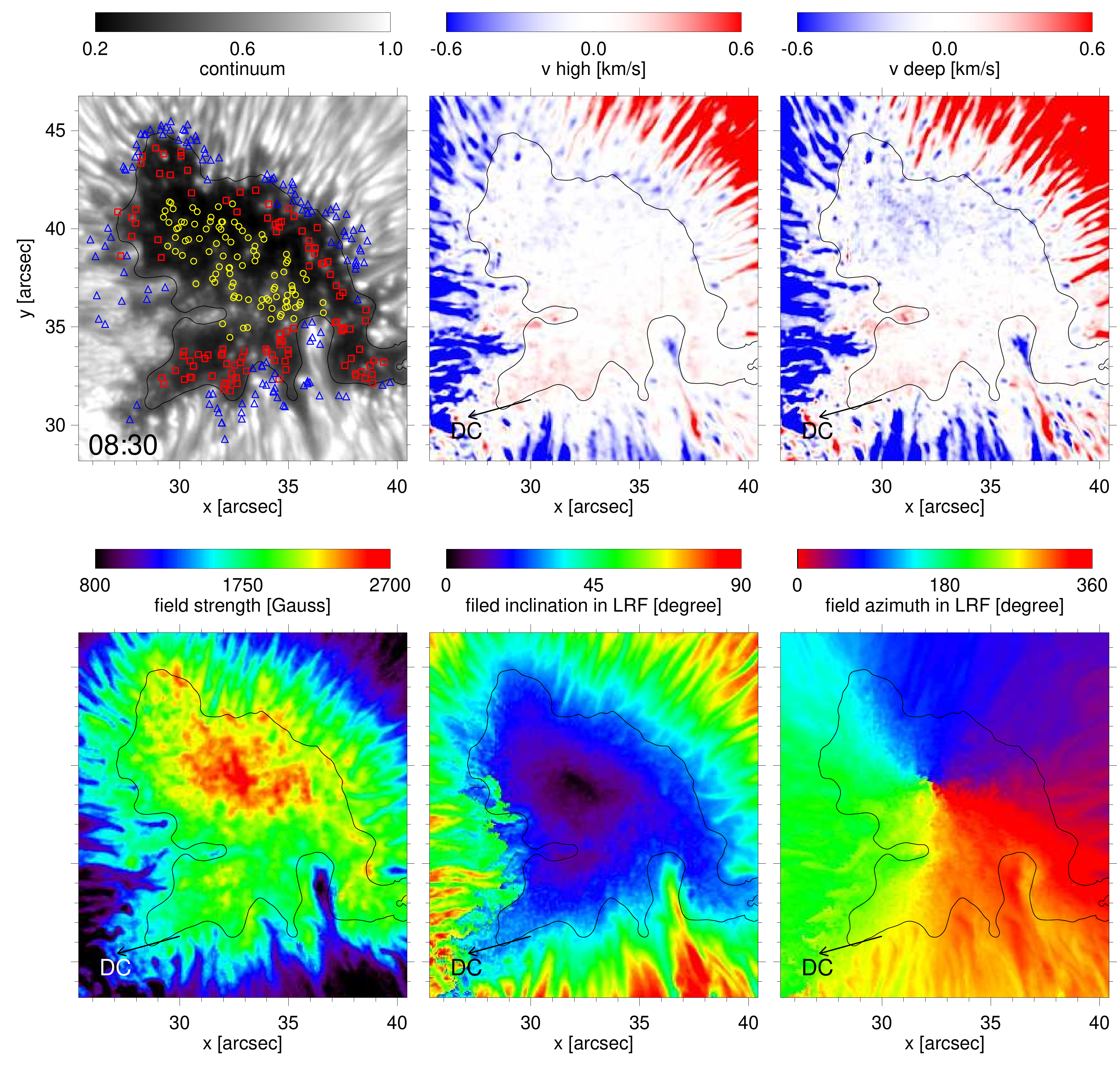}}
\caption{Physical parameters derived from the observed Stokes
  profiles. Shown in the figure is the scan with best seeing
  conditions, taken at 08:30\,UT.  The top panels, from left to right,
  display the continuum intensity map with the position of appearance
  of all the detected \uds\ (yellow circles are central \uds, red
  squares are peripheral \uds, blue triangles are grain-origin \uds),
  and two bisector maps sampling high (\vhigh) and deep (\vdeep)
  photospheric layers.  Negative velocities (blue) represent upflows
  along the line of sight, while positive velocities (red) mean
  downflows.  The bottom panels, from left to right, show the field
  strength, the field inclination, and the field azimuth in the local
  reference frame (LRF).  The black contours indicate the approximate
  umbral boundary defined by the spatially-smoothed continuum
  intensity.  The arrows mark the direction to disk center (DC). An
  mpeg version of this figure is available in the on-line journal.
  } \label{fig:data_reduction}
\end{figure*}

Figure~\ref{fig:cut_image} displays a close-up of the umbra and the
inner penumbra on the disk-center side of the spot. The maps of mean
linear polarization (LP) and circular polarization (CP) degree were
calculated as
\begin{equation}
{\rm LP}=\frac{\int{[Q^{2}(\lambda) + U^{2}(\lambda)]^{1/2}\,/\,I(\lambda)}\, {\rm d}\lambda}{\int{{\rm d}\lambda}}, 
\end{equation}
\begin{equation}
{\rm CP}=\frac{\int{|V(\lambda)|\,/\,I(\lambda)}\,{\rm d}\lambda}{\int{d\lambda}} ,
\end{equation}
with the integration extending over the \ion{Fe}{1} 6301.5~\AA\/ line. 
The LP signals are stronger toward the limb due to projection
effects.  Most of the fine-scale structures seen in the LP and CP maps
coincide with \uds.  Figure~\ref{fig:cut_image} clearly demonstrates
that
\begin{enumerate}
\item The first and last intensity images of the scan are perfectly
  aligned and do not show distortions in spite of the fact that they
  were taken about 30\,s apart (top row). This confirms the good
  performance of the MOMFBD processing.
\item Almost all \uds\ in the umbra keep their brightness, size, and
  position during the line scan (top row), that is, the scan time is
  shorter than the dynamic timescale of \uds.
\item The smallest features visible in the maps (\uds\ and dark cores of
  penumbral filaments) are only slightly larger than the diffraction
  limit of the SST.  They show higher contrasts in circular polarization.
\end{enumerate}
We believe this is one of the best \ud\ data sets ever obtained,
because of its superb spatial resolution (0\farcs13), the stable
seeing conditions, and the availability of two-dimensional
spectropolarimetric measurements for more than 70 minutes with good
temporal cadence (63\,s).  The non-simultaneous acquisition of
spectral information---the main disadvantage of Fabry-P\'erot
systems---can be well neglected because
\uds\ evolve on time scales longer than the scan time.


\section{Data analysis}\label{sec:reduction}

In this section we describe the methods we have used to detect and
track \uds, as well as the line bisector calculations and Stokes
inversions performed to derive their velocities and magnetic fields.

\subsection{Detection and Categorization}\label{sec:detection}

For the statistical study of \uds\ it is convenient to use automatic
detection algorithms \citep{1997A&A...328..682S, 2010A&A...510A..12B}.
However, we have implemented a manual procedure here because, even
under the very stable seeing conditions of our observations, the image
quality still shows an unavoidable amount of residual fluctuation
which might compromise the performance of automatic methods.

The procedure works as follows.  First we inspect the continuum movie
to identify the frame of appearance of each \ud. The position of the
\ud\ is then tracked by clicking on the screen until it disappears.
After going through the temporal sequence, we run the movie backward
in time to check the consistency.  When an \ud\ splits in two
fragments, the bigger component is the one that continues to be
tracked and the smaller component is selected separately as a new
entity.  When two \uds\ merge, the smaller is assumed 
to die.  In some cases, \uds\ show recurrence at the same position.  
If the recurrent \uds\ appear within an interval of 3 frames
($\sim$3\,min), we consider them a single entity.

\uds\ evolving from penumbral grains are also studied.  In this case,
the detection starts from the frame in which the tip of the penumbral
grain detaches from the filamentary structure.  Sometimes these \uds\ 
are connected to the penumbral grains through a faint tail, but they 
become more isolated and roundish as they move toward the umbra.

This method has allowed us to obtain the trajectories of 339
\uds.  According to their place of birth, we categorize
them into three groups. \Uds\ located in the central part of the umbra
are called ``central \uds'' (98 samples out of 339 \uds), 
\uds\ located in the peripheral area are called ``peripheral \uds'' 
(112 samples), and \uds\ detached from penumbral grains are called
``grain-origin \uds'' (129 samples).  The peripheral area is an
$\sim$1--2\,\arcsec\ annular region adjacent to the umbra-penumbra
boundary.  It is known that central \uds\ are static while peripheral
and grain-origin \uds\ show a systematic motion toward the center of
the umbra \citep{1992SoPh..137..215E, 1997A&A...328..689S}. 
About 40\%\ of the 339 \uds\ we have detected did not appear or
disappear within the interval covered by the observations. Therefore, 
their lifetimes could not be computed. 

\begin{table}[t]
\begin{center}
\caption[]{\label{tbl:table1} Average properties of \uds}
\begin{tabular}{lrrr}
\tableline 
Parameter            & Central & Peripheral & Grain-origin \\
\tableline
Number				   & 98 & 112 & 129 \\
Lifetime{\footnotemark[1]} [min]   & 19 & 18 & 17 \\
Proper motion [\kms]   & 0.19 & 0.31 & 0.49 \\
Brightness ratio       & 1.46 & 1.50 & 1.89 \\  
Diameter [km]          & 401 & 391 & 482 \\   
\tableline
\end{tabular}
\end{center}
\footnotetext[1]{Excluding \uds\ which do not appear or disappear within the 
observation period.}
\end{table}

\begin{figure}
 \centerline{\includegraphics[width=90mm, bb=0 0 680 510]{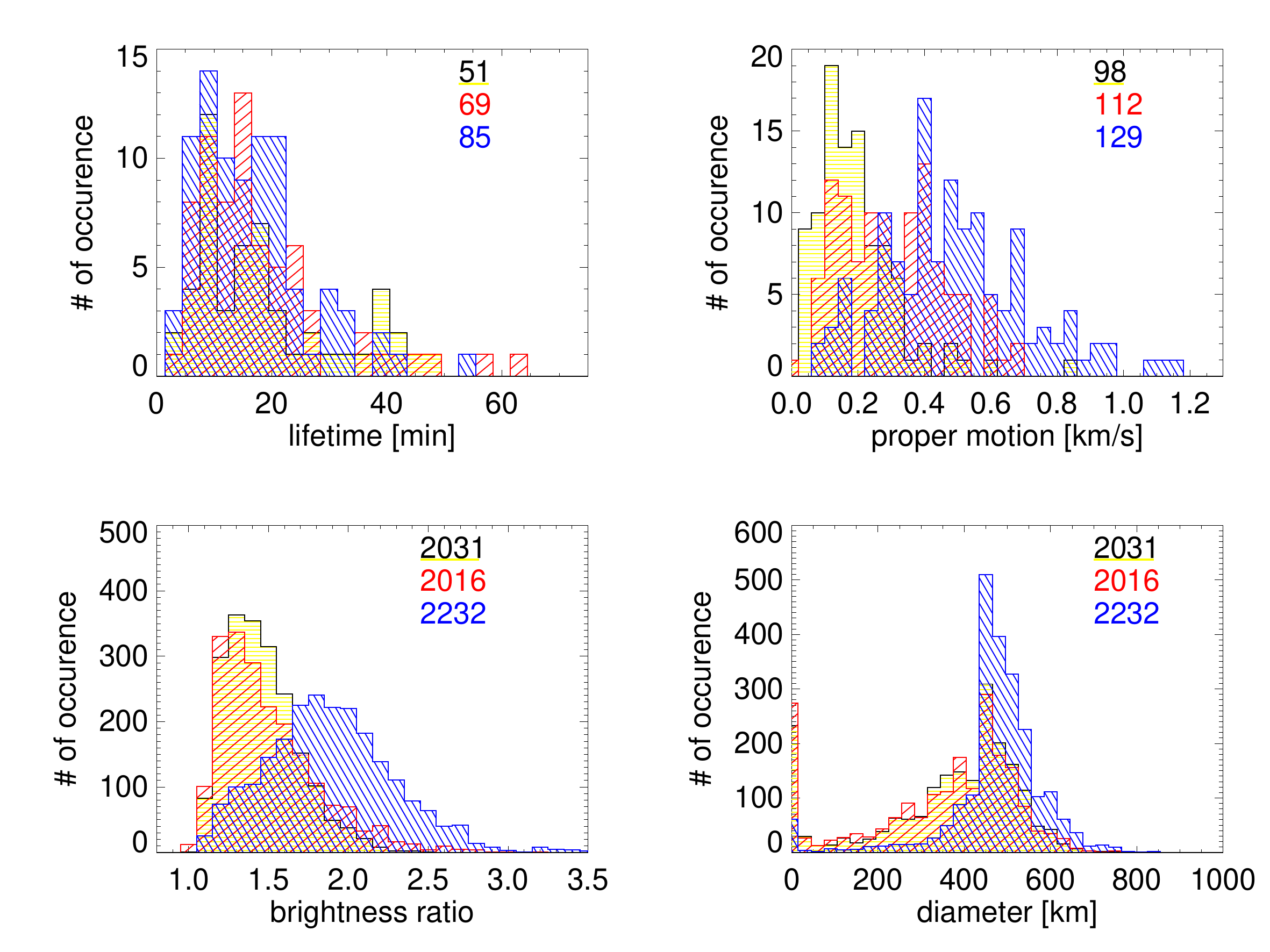}}
 \caption{Histograms of UD lifetime (upper left), average proper
 motion speed (upper right), brightness ratio (lower left), and
 diameter (lower right).  Central \uds\ are displayed with black bars
 filled with yellow color, peripheral \uds\ with red color, and
 grain-origin \uds\ with blue color. The number of measurements in
 each category is shown in the upper-right corner of the histograms. }
 \label{fig:histogram_4parameter}
\end{figure}

In the upper left panel of Figure~\ref{fig:data_reduction} we show the
position of appearance of all 339 \uds.  The yellow circles indicate
central \uds, the red squares peripheral \uds, and the blue triangles
grain-origin \uds.  Table~\ref{tbl:table1} lists their mean
properties. The average lifetime is about 18~min.  This is relatively
long compared with the values reported in previous works
\citep[e.g.,][]{2008A&A...492..233R}, probably because our manual
procedure is capable of detecting fainter \uds.  UDs move with an
average speed of 0.2-0.5\,\kms. The proper motion speed is
defined as the distance between the points of appearance and
disappearance divided by the time interval.  Distances are not
corrected for projection effects. The brightness ratio is the
\ud\ intensity (the maximum intensity within a $\pm$2~pixel area)
relative to the intensity of the dark background ($I_{\rm db}$). We
use the ``dark background''---the region surrounding the UD but
excluding the UD itself---as a local reference. $I_{\rm db}$ is the
mean intensity of pixels over a $2\arcsec \times 2\arcsec$ area
centered in the \ud\ whose intensity is darker than the average minus
0.5$\sigma$. Here, $\sigma$ represents the standard deviation of the
intensity within the $2\arcsec \times 2\arcsec$ area. We find average brightness ratios from 1.5
(central UDs) to nearly 1.9 (grain-origin UDs). The mean \ud\ diameter
is 400--500 km. For each UD, the diameter is calculated as the average
distance in eight radial directions along which the intensity is
brighter than $1.2 \, I_{\rm db}$, or as the distance to the closest
inflection point.  A similar method was adopted in
\citet{2009ApJ...702.1048W}.  There are some instances of zero
diameter, which means that the \ud\ intensity was darker than $1.2 \,
I_{\rm db}$.  Brightness ratios and diameters are calculated for
every time step within the \ud\ lifetime, so in total we get 6279
individual values.

\begin{figure}[t]
 \centerline{\includegraphics[width=75mm, bb=0 0 510 453]{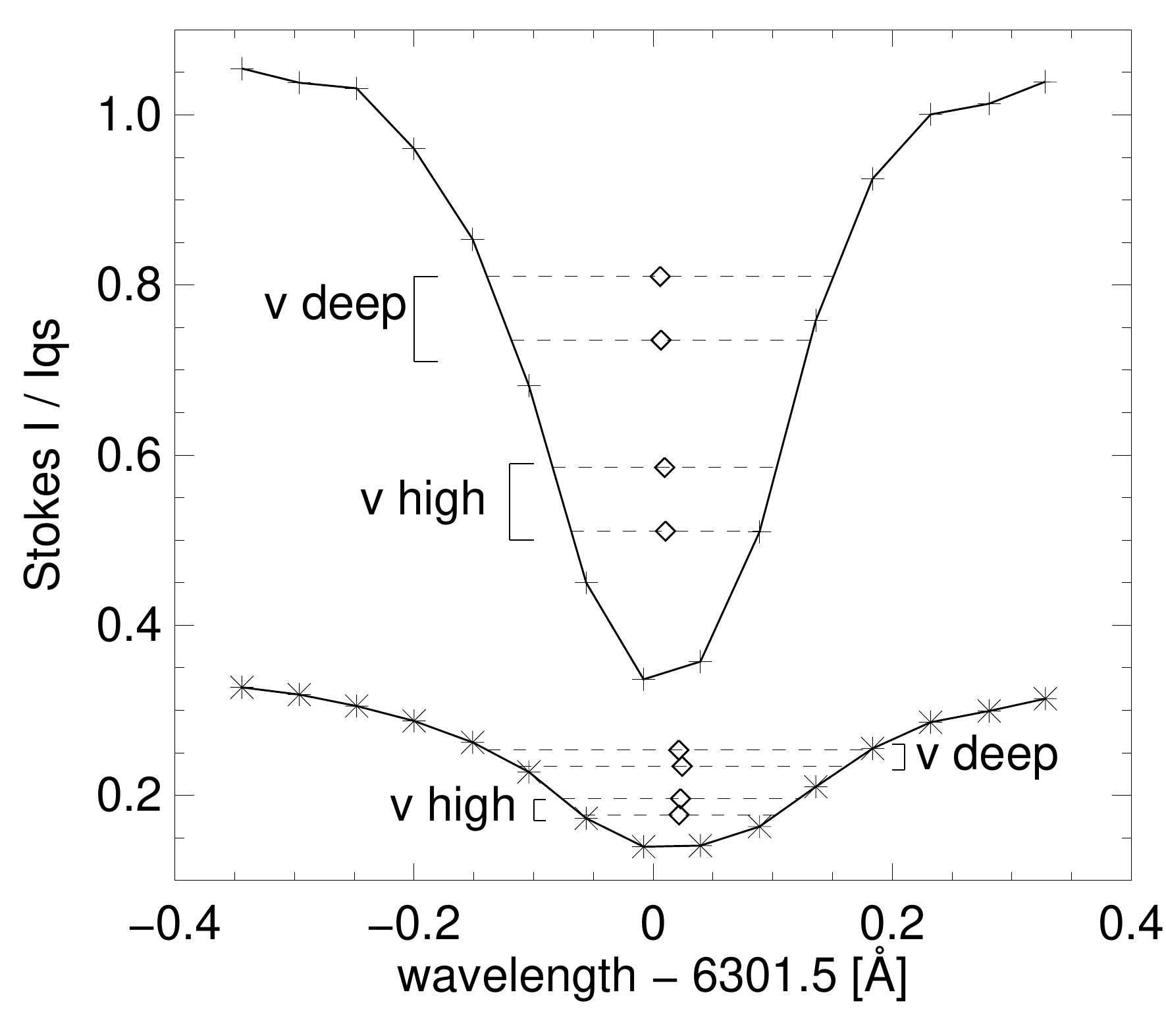}}
 \caption{Intensity profiles emerging from the tip of a penumbral
 grain (pluses) and from an \ud\ (asterisks).  The diamonds indicate
 the bisector positions at intensity levels of 10\% and 23\% (constituting \vhigh),
 and 49\% and 62\% (constituting \vdeep). } \vspace*{1em}  \label{fig:bisector_profile}
\end{figure}

Figure~\ref{fig:histogram_4parameter} shows histograms of these
parameters for all the UDs detected in the observations. Central and
peripheral UDs are very similar except that the latter move
faster. Grain-origin UDs, however, are clearly different: they have
the fastest speeds, the highest intensity contrasts, and the largest
diameters.

\subsection{Derivation of Velocities: Line Bisectors}\label{sec:bisector}

\begin{figure}[t]
\centerline{\includegraphics[width=82mm, bb=0 0 340 708]{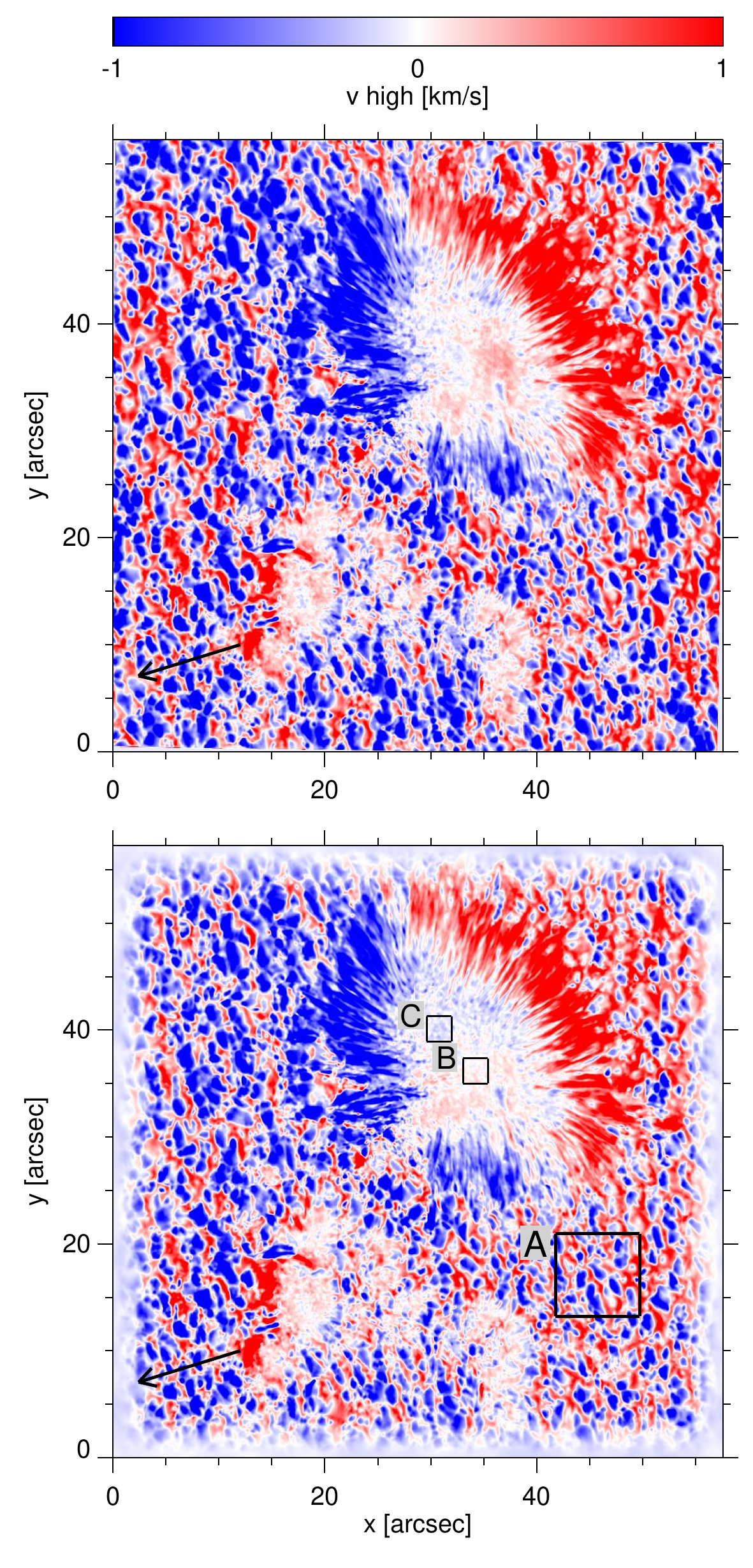}}
\caption{Original \vhigh\ map observed at 08:31\,UT (top) and
  subsonic filtered map (bottom).  Negative velocities (blue) mean
  upflows along the line of sight, and positive velocities (red) mean
  downflows.  The arrows mark the direction to disk center. The
  regions labeled A--C will be studied in Section~\ref{sec:correlation}.}  
  \label{fig:subsonic_filter}
\end{figure}

\begin{figure}[t]
\centerline{\includegraphics[width=85mm, bb=0 0 340 255]{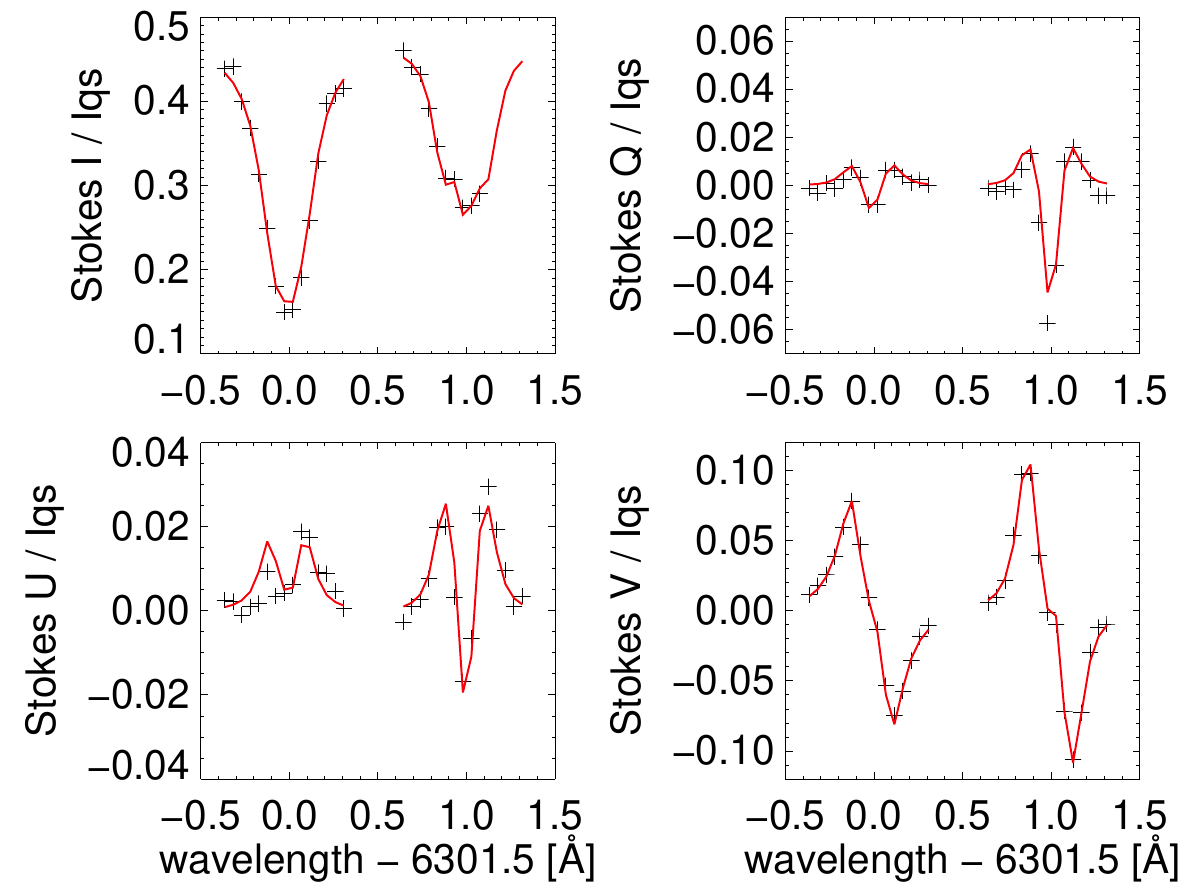}}
   \caption{Stokes profiles observed at the position of an \ud\
   (marked with a triangle in Figure~\ref{fig:fig1}) and best-fit
   profiles returned by SIR (symbols and red solid lines,
   respectively).  \vspace*{1em} } \label{fig:sir_profile}
\end{figure}

The mean wavelength position of the two line wings at a given
intensity level is called the line bisector
(Figure~\ref{fig:bisector_profile}).  Since different intensity levels
sample different atmospheric layers, bisectors are often used to
estimate the height variation of the line-of-sight velocity
\citep[e.g.,][]{2004A&A...415..717T, 2010ApJ...713.1282O}.  The 0\%
intensity level represents the line core, while 100\% means the local
continuum.

We calculate bisectors only for \ion{Fe}{1}\ 6301.5\,\AA\ because the
red wing of \ion{Fe}{1}\ 6302.5\,\AA\ is strongly blended with the
telluric O$_{2}$ 6302.8\,\AA\ line (see the Stokes $I$ profile in
Figure~\ref{fig:fig1}).  The \ion{Fe}{1}\ 6301.5\,\AA\ line is less
affected by blending, although its far blue wing (intensity levels $>
70\%$) sometimes show influence of molecular lines in the darkest
umbral areas \citep{1993A&A...270..494M}, which may result in a
systematic blueshift. For this reason we calculated four bisector
positions at intensity levels of 10\%, 23\%, 49\%, and 62\%, avoiding
bisectors close to the continuum.  Each bisector position is obtained
from a linear interpolation of the relevant intensities in the
observed line profile.  To reduce the noise, two bisector levels are
averaged.  As can be seen in Figure~\ref{fig:bisector_profile}, this
results in two bisector velocities which will be called \vhigh\ (10\%\
and 23\%) and \vdeep\ (49\%\ and 62\%).  The \vhigh\ and \vdeep\ maps
are not corrected for projection effects, but they mostly represent
vertical flows because horizontal flows are usually weak in the solar
photosphere. Negative values of \vhigh\ and \vdeep\ correspond to
blueshifts (or upflows along the line of sight), and positive values
to redshifts (or downflows).

Following \citet{1989ApJ...336..475T}, we apply a subsonic Fourier
filter to the temporal sequence of \vhigh\ and \vdeep\ to suppress
disturbances with horizontal speeds larger than 4\,\kms, which are
mostly due to $p$-modes and the residual temporal noise.  In the
umbra, a large scale ($\sim$5000\,km) velocity pattern corresponding
to the $p$-mode oscillations is observed. This pattern is removed by
the filter. We use an edge apodization of 10\%\ in space and time to
avoid the propagation of boundary errors.  The effect of the subsonic
filtering is demonstrated in Figure~\ref{fig:subsonic_filter}. The
dominant redshift signal on the right side of the umbra at $(x, y)
\sim (37\arcsec,37\arcsec)$ in the original image (upper panel) is
absent in the filtered image (lower panel).  However small local
variations, which coincide with the \ud's positions, keep their
identity even after the subsonic filtering.

For both \vdeep\ and \vhigh, the zero velocity is determined by
averaging the filtered \vhigh\ map over umbral pixels with intensities
below $0.4\,I_{\rm qs}$.  This is done for each of the 68 frames of
the sequence.  The \vhigh\ and \vdeep\ maps look similar, but the
latter shows larger root-mean-square (rms) fluctuations. The typical
rms velocities in the umbra are 0.11\,\kms\ for
\vhigh\ and 0.13\,\kms\ for \vdeep. 
The rms variations increase significantly when the maps are not
filtered. Thus, the filtering is essential to provide a good velocity
reference, and also to remove the large velocity offsets induced by
the $p$-mode oscillations. This is particularly important when dealing
with small velocities such as those observed in \uds.

\subsection{Derivation of Magnetic Properties: Stokes 
Inversion}\label{magnetic_inversion}

\begin{figure*}[t]
\centerline{\includegraphics[width=180mm, bb=0 0 793 552]{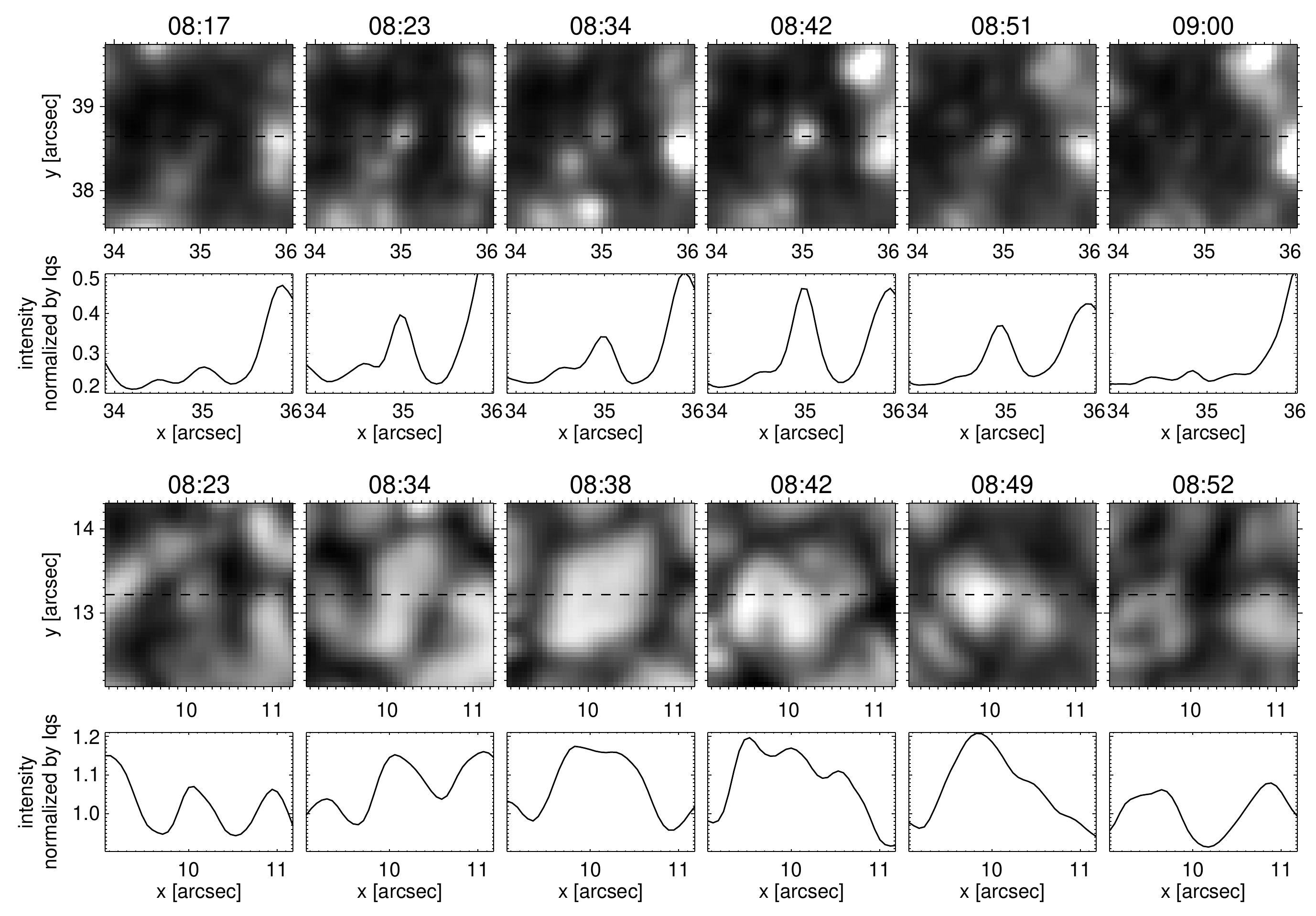}}
   \caption{Temporal evolution of a central \ud\ (top) and a quiet Sun
   granule (bottom). Displayed are continuum intensity maps for an
   area of $2\farcs2 \times 2\farcs2$.  The coordinate system
   coincides with that of Figure~\ref{fig:fig1}. An intensity cut
   along the dashed line plotted at the center of the FOV is shown
   below each image.  Times are given in UT in the intensity
   panels.} \label{fig:convection_morphology}
\end{figure*}

We determine the magnetic properties of the umbra by inverting the
observed Stokes profiles with the SIR code \citep[Stokes Inversion
based on Response functions;][]{1992ApJ...398..375R}. The two lines
are fitted simultaneously, excluding the telluric O$_2$ blend in the
red wing of the \ion{Fe}{1} 6302.5 Stokes $I$ profile.

The inversion is carried out in terms of a one-component model
atmosphere with constant (i.e., height-independent) magnetic fields
and velocities, which is sufficient to explain the relatively
symmetric Stokes profiles observed in the sunspot umbra (see
Figure~\ref{fig:sir_profile}). Zero stray-light contamination and
unity magnetic filling factors are assumed. The inversion returns 9
free parameters: the three components of the vector magnetic field
(strength, inclination, and azimuth), the line-of-sight velocity, the
microturbulent velocity, and the temperature at 4 nodes.  The initial
guess model used to start the inversion is the hot umbral model of
\citet{1994A&A...291..622C}.

Sample maps of the retrieved magnetic parameters are displayed in
Figure~\ref{fig:data_reduction}.  The inclination and azimuth angles
are expressed in the local reference frame to avoid projection
effects. The inclination varies from $0^\circ$ to $180^\circ$ for
vertical fields pointing away from and to the solar surface,
respectively. The azimuth is measured counterclockwise from the
positive $x$-axis of the figure. As expected, the sunspot shows an
outward-directed radial magnetic field.  We also note that the
line-of-sight velocities returned by the inversion agree well with the
bisector velocities obtained from the 6301.5\,\AA\ line.  An animated
version of Figure~\ref{fig:data_reduction} is available in the on-line
journal.


\section{Convection in the Umbra}\label{sec:convection}

In the solar photosphere, the presence of strong magnetic fields
inhibits convective energy transport. This is why sunspots are dark
and cool compared to the quiet Sun, which is covered by convective
cells called granules. However, convection is not entirely suppressed
even in the umbra: as a matter of fact, \uds\ are the manifestation of a
modified convective pattern. The properties of this pattern still need
to be determined. Here we study how the characteristics of convection
differ between the quiet Sun and the umbra, paying special attention
to the morphology of the convective cells and the correlation between
brightness and velocity.

\subsection{Morphology}\label{sec:morphology}
 
Figure~\ref{fig:convection_morphology} illustrates the morphological
evolution of a central \ud\ and a quiet-Sun granule.  The images cover
an area of $2\farcs2 \times 2\farcs2$.  First we notice that the
\ud\ is much smaller than the granule \citep[$\sim$0\farcs3 vs
1\farcs0; cf.,][]{1987SoPh..107...11R}.  Second, \uds\ are relatively
isolated whereas granules are closely packed between narrow
($\sim$0\farcs3) intergranular lanes.  Third, the intensity profile of
granules is flat-top, sometimes with intensity depressions in the
middle.  A granule loses its identity (or disappears) by fragmentation
or by merging with neighboring granules.  On the other hand, the
intensity profile of
\uds\ has a Gaussian shape, and they disappear mostly by fading out.

Both granules and UDs exhibit a turbulent character: the structures
displayed in Figure~\ref{fig:convection_morphology} change their
brightness, barycenter, and shapes on timescales of only a few
minutes. The variations are more pronounced in the case of
granules. The average lifetime of UDs, 18 minutes, is slightly 
longer than the 5--15 minute duration of granules
\citep[e.g.,][]{1961ApJ...134..312B, 1987A&A...174..275A,
1989ApJ...336..475T,1999ApJ...515..441H}.

\begin{figure}[t]
\centerline{\includegraphics[width=85mm, bb=0 0 340 481]{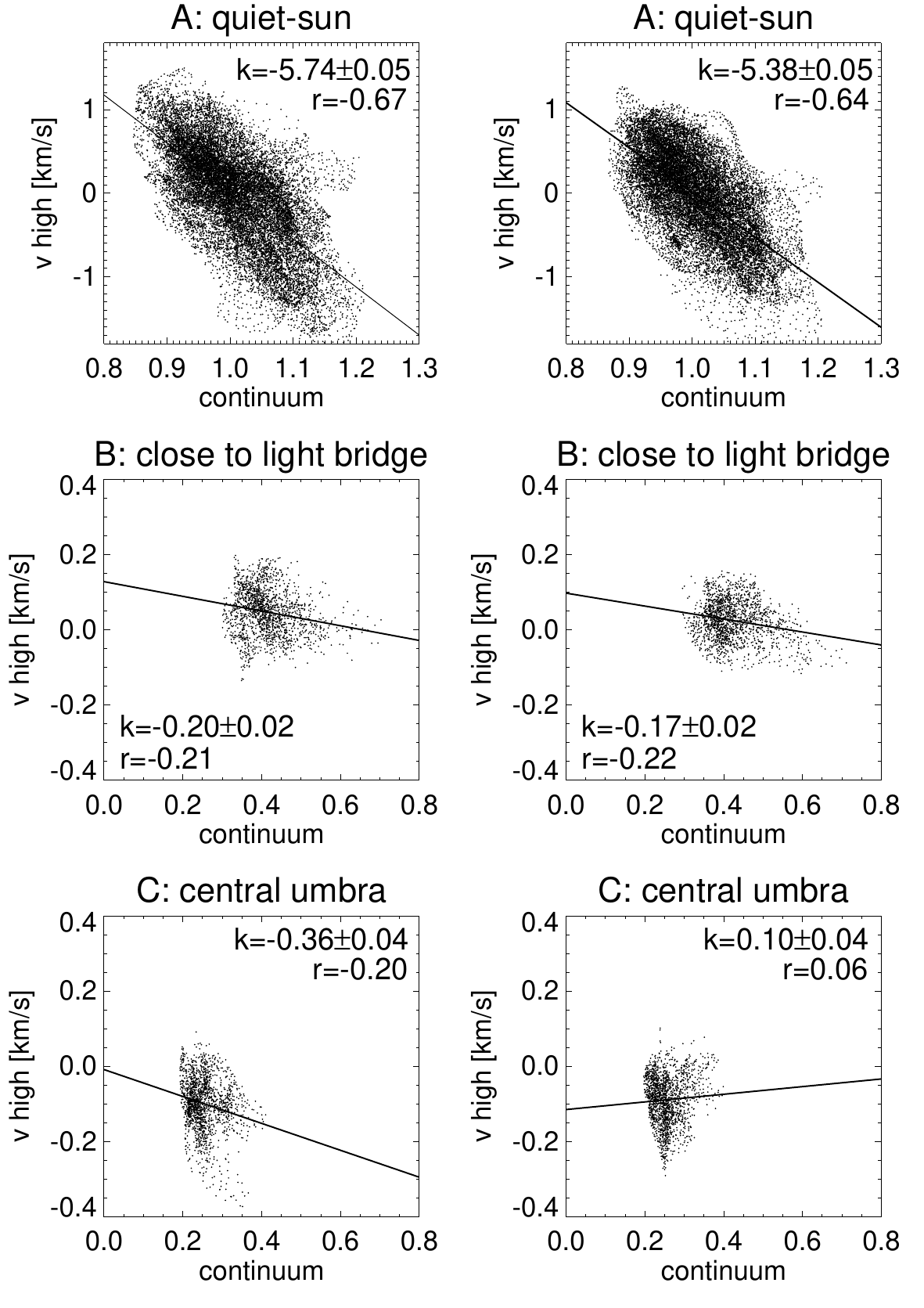}}
\caption{Bisector velocity \vhigh\ vs continuum intensity for 
selected umbral and quiet Sun areas using the line scans taken at
08:28\,UT (left column) and 08:33\,UT (right column). Negative
velocities represent upflows along the line of sight.  The positions
of areas A--C are indicated in Figure~\ref{fig:subsonic_filter}.  $k$
is the slope of the best linear fit and $r$ the Pearson
correlation coefficient.}
\vspace*{1em}
\label{fig:bisector_scatter}
\end{figure}

\subsection{Brightness vs Velocity}\label{sec:correlation} 

The strong correlation between brightness and line-of-sight velocity
in granular convection is well known.  Bright areas (i.e., granules)
show upflows, while dark areas (the intergranular lanes outlining the
granules) harbor downflows.  This is illustrated in the top panels of
Figure~\ref{fig:bisector_scatter}.  

In the umbra, in addition to \uds\/, there exist diffuse areas with
enhanced brightness. These areas may correspond to a convective
pattern similar to (but weaker than) that of the quiet Sun, with \uds\
being another manifestation of the same pattern occurring at positions
where convection is more vigorous. To test this possibility, we have
examined the correlation between brightness and velocity in the umbra:
if bright structures are the result of convection, then they should
preferentially be associated with upflows.  The middle and bottom
panels of Figure~\ref{fig:bisector_scatter} show scatter plots for two
umbral areas labeled B and C in Figure~\ref{fig:subsonic_filter}.
Area~B is close to the light bridge and C represents the central
umbra. In both areas, at 08:28 UT (left column) we find a tendency of
upflows in bright regions and downflows in dark regions similar to
that of the quiet Sun, yet with weaker correlation. For most of the
scans with good seeing conditions, the slope $k$ of the best linear
fit to the velocity-brightness relation turns out to be negative, with
average values of $-0.11 \pm 0.02$ in area~B and $-0.24 \pm 0.04$ in
area~C. However, the negative correlation does not always persist in
time and scans obtained a few minutes apart under good seeing conditions 
sometimes show opposite behaviors (compare the lower panels of
Figure~\ref{fig:bisector_scatter}). For this reason, the present data
do not allow us to unambiguously confirm the existence of a global
convective pattern in the umbra.  We will see later that the situation
is different for \uds.

\begin{figure}[t]
\centerline{\includegraphics[width=85mm, bb=0 0 340 382]{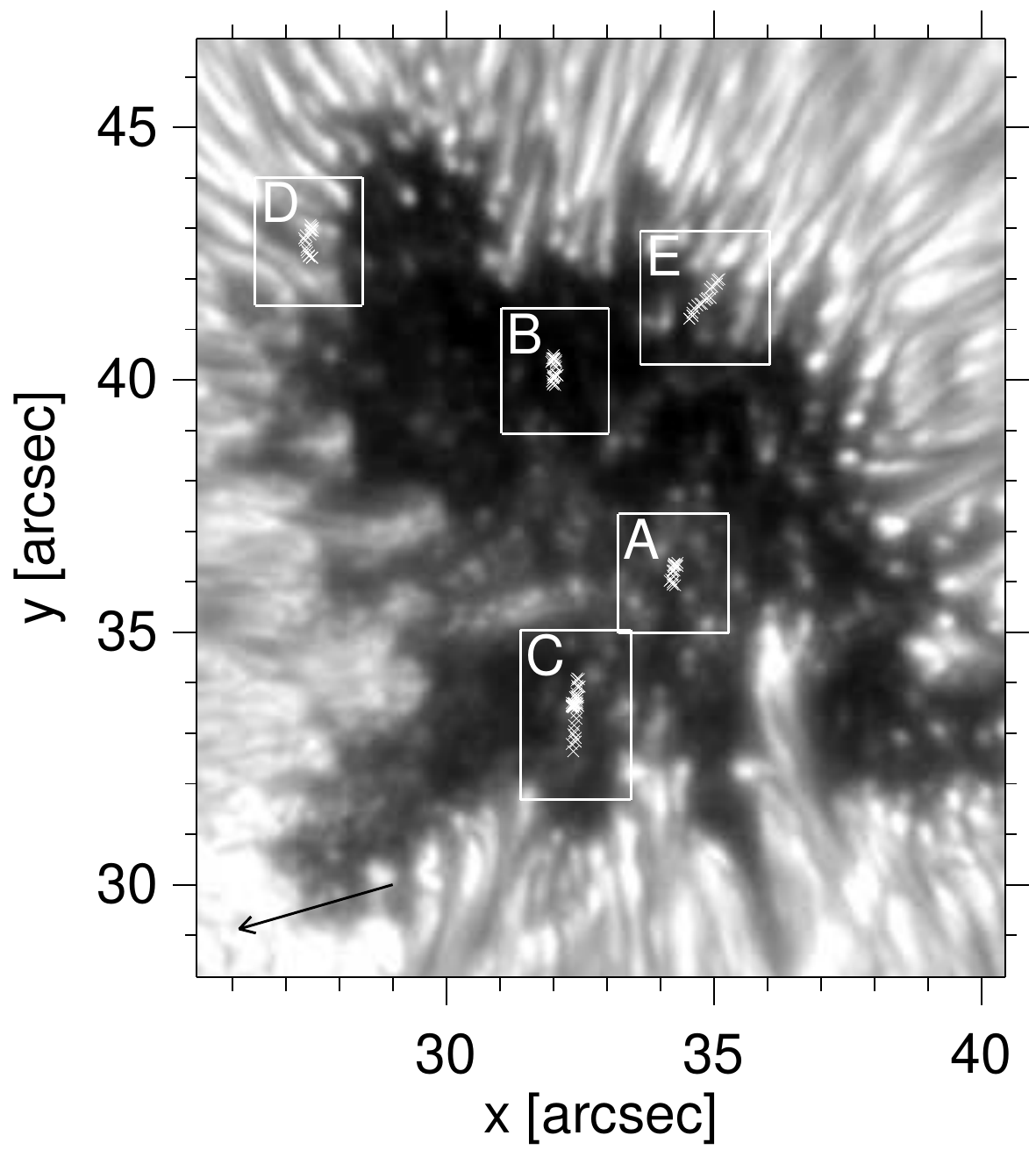}}
   \caption{Position of selected \uds\ (A-B: central UDs, C:
   peripheral \ud, D-E: grain-origin UDs). The background image shows
   the continuum intensity at 08:30\,UT. The cross symbols indicate
   the trajectories of the \uds. The arrow marks the 
	direction to disk center. }	
	\vspace*{2em}  \label{fig:casestudy_position}
\end{figure}


\section{Umbral Dots}\label{sec:UDs}

In this Section we give a detailed description of the 
evolution and statistical properties of individual \uds. 

\subsection{Case Studies}\label{sec:case_study}

We select five \uds\ (UD\#A--E) whose locations are indicated in
Figure~\ref{fig:casestudy_position}.  All of them were observed from
appearance to disappearance. Movies of their temporal evolution can
be found in the electronic journal.

In Section~\ref{sec:verydeep}, another two \uds\ from one of the best
scans of the sequence will be considered. We use them to demonstrate
the existence of localized downflow patches around \uds\ in deep
photospheric layers.

\subsubsection{Typical Central \ud}

According to our visual inspection, more than 70\% of the central \uds\ 
do not show flow field perturbations, i.e., no upflows or downflows
are detected.  Similarly, magnetic field perturbations associated with
central \uds\ are usually not visible or very small.  17\% of the central
\uds\ split or merge with neighboring UDs. 

The temporal evolution of a typical central \ud\ (UD\#A) is displayed
in Figures~\ref{fig:centralUD_34_sequence} and
\ref{fig:centralUD_34_plot}.  Figure~\ref{fig:centralUD_34_sequence}
shows maps of physical parameters starting 2 frames before the
appearance of the UD (T=0\,s) and ending 2 frames after its
disappearance.  In the left column of
Figure~\ref{fig:centralUD_34_plot}, curves representing the temporal
variation of the parameters at the position of the UD (solid line) and
the dark background (dashed line) are given.  The curve for the dark
background is an average over pixels with intensities lower than the
average minus $0.5\sigma$ in the FOV of
Figure~\ref{fig:centralUD_34_sequence}. The right column of
Figure~\ref{fig:centralUD_34_plot} shows spatial profiles along the
$x$-direction at $y$-positions co-moving with the
\ud.  The Gaussian peak at $x = 34\farcs3$ in the continuum
plot corresponds to the \ud.

UD\#A was born in a diffuse bright region, then increased in
brightness (0\,s$<$T$<$377\,s), and finally merged with a neighboring
\ud\ (755\,s$<$T$<$1006\,s).  As can be seen in both maps and plots,
it did not show clear upflows or downflows.  During most of its
lifetime, the UD was located in a patch of enhanced redshifts whose
morphology and amplitude do not correlate with the UD evolution.
Other UDs in the FOV display localized upflows of up to
\vdeep\ $=$ 0.3\,\kms, but they do not persist in time. An
example is the structure located next to UD\#A at coordinates
(34\farcs1, 36\farcs6), between $T=566$ and 755~s.  In
Figure~\ref{fig:centralUD_34_sequence}, a local reduction of the field
strength at the position of UD\#A can be seen during
0\,s$<$T$<$629\,s. Then, the \ud\ appears to gradually merge with a
pre-existing weak field patch right above it.  The amplitude of the
magnetic field perturbations is very small, about 50~G. The UD does
not leave clear signatures in the inclination maps.

\subsubsection{Distinct Central \ud}

Figures~\ref{fig:centralUD_42_sequence} and
\ref{fig:centralUD_42_plot} show the temporal evolution of a distinct
central \ud\ (UD\#B).  This \ud\ differs from the others because of
its large brightness and upflow.  UD\#B was born in a very dark area
within the umbra and the continuum intensity became quite high about 9
minutes later (566\,s$<$T$<$880\,s).  A second smaller intensity peak
appeared in the interval 1069\,s$<$T$<$1321\,s.  Finally the UD faded
in a diffuse bright background with no detectable intensity peak.
There was a significant upflow associated with the continuum intensity
enhancement.  The maximum upflow of \vhigh\
$\sim 0.3$~\kms\ occurred at T=755\,s.  A reduction of the field
strength of the order of 50~G co-spatial with the \ud\ can be observed
during the first half of its lifetime.  After T=755\,s, the region of
weaker field strengths disappears and the \ud\ collides with a
pre-existing strong field region.  Sometimes there is a small spatial
displacement (up to 0\farcs2) between the brightness peak and the
patch of reduced field strengths.  A very small patch with more
inclined fields appeared transiently from $T=566$ to 755~s.

The bright UD located at coordinates $(31\farcs5, 40\arcsec)$ also
showed reduced field strengths and strong upflows (particularly in
\vhigh) for more than 6 minutes, from $T = -125$ to 251~s.

\subsubsection{Typical Peripheral \ud}

Peripheral \uds\ are born in the peripheral region of the umbra, where
the continuum intensity is brighter and the magnetic field more
inclined.  A significant property of peripheral \uds\ is their
systematic motion toward the center of the umbra (see Table~1). 
Our visual inspection reveals that 55\% of the detected peripheral
\uds\ show upflows, but no systematic magnetic field
perturbations. For example, 13\% of peripheral \uds\ present a
reduction of field strength, while 16\% of them are associated with
enhanced fields. Usually, these perturbations do not last more 
than a few minutes.

Figures~\ref{fig:PUD_036_sequence} and \ref{fig:PUD_036_plot} display
the evolution of UD\#C, a typical peripheral UD. The inward migration
of this \ud\ is very clearly seen in the online animation.  The
migration speed is $\sim$1.1\,\kms\ during 0\,s$<$T$<$503\,s, and
almost zero during 503\,s$<$T$<$1635\,s.  In the final stages of its
life (1635\,s$<$T$<$2390\,s), the UD migrates again with a speed of
0.5\,\kms.  The origin of UD\#C is a diffuse bright area within the
umbra. It shows three brightness peaks: at $T=377$, 1132, and 1761~s.
Only the first is associated with strong upflows (\vhigh\
$\sim -0.15$~\kms\ and \vdeep\ $\sim -0.35$~\kms). The other two
peaks show much smaller velocities, but still shifted to the blue
compared with the dark background. A hint of redshifts can be seen in
\vdeep\ from T$=$2202 to 2390\,s.

\begin{figure*}[t]
\centerline{\includegraphics[width=170mm, bb=0 0 623 364]{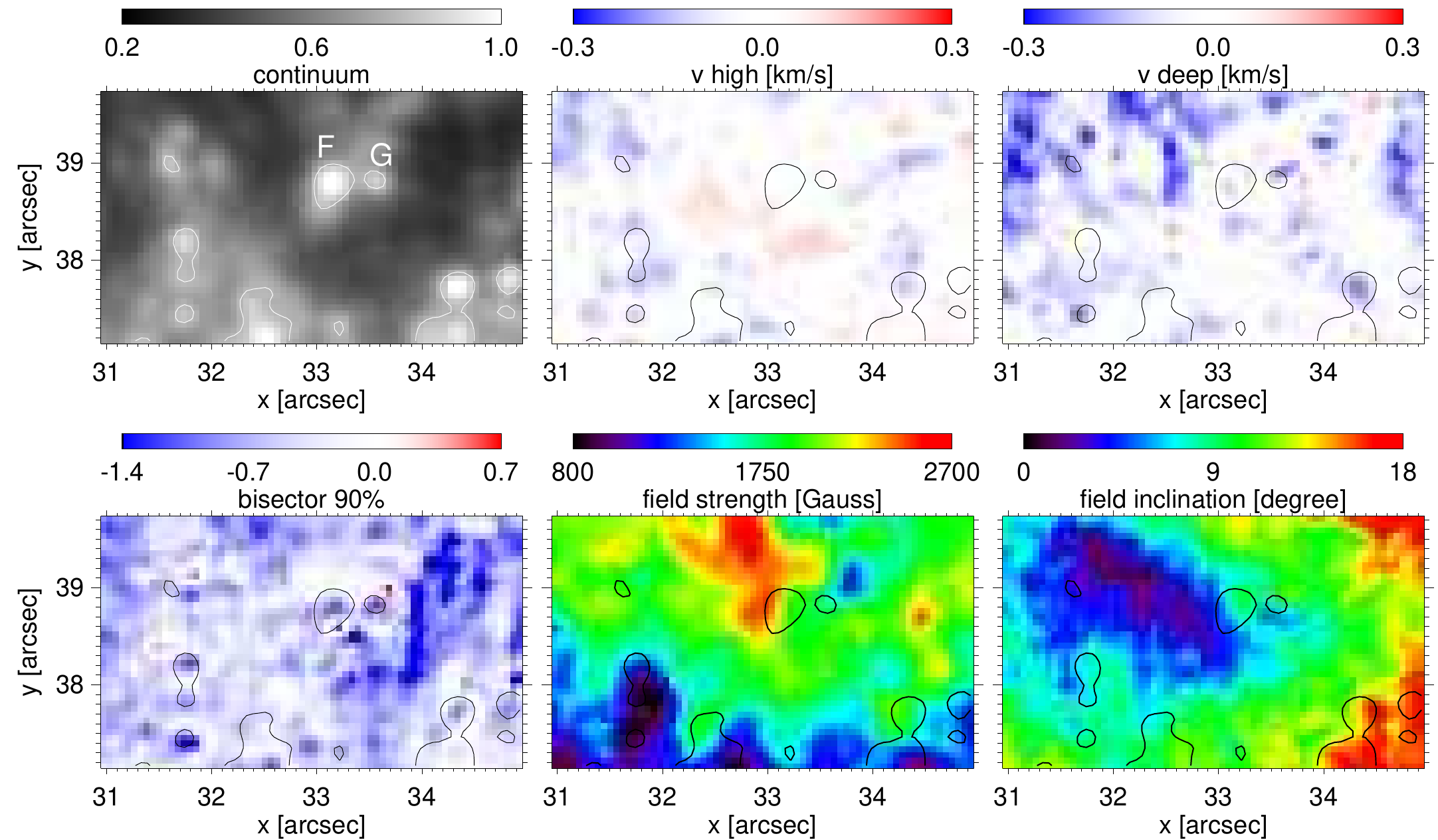}}
   \caption{Close-up images of the central area of the umbra from the
   line scan taken at 08:30\,UT. {\em Clockwise, starting from top
   left:} continuum intensity, \vhigh, \vdeep, field inclination,
   field strength, and bisector velocity at the 90\% intensity
   level. Negative velocities represent upflows along the line of
   sight.  The contours show continuum intensities of $0.4 \, I_{\rm
   qs}$.  } \label{fig:verydeep}
\end{figure*}

The magnetic field shows a complex distribution. A patch of reduced
field strength ($\sim -100$~G) develops at the position of the UD and
migrates inward to the umbra during 377\,s$<$T$<$1069\,s. The magnetic
field perturbation then disappears. At T$=$1635\,s the negative patch
can be observed again at the position of the UD.  It will survive for
most of the remaining UD's evolution. In addition, another patch of
increased field strengths is visible to the left of the UD, at
$(x,y)\approx (32\arcsec, 33\arcsec)$.  This structure already exists from
T=$-$125\,s (before the appearance of UD\#C) and also migrates inward,
but with slower speed (0.5\,\kms\ during 0\,s$<$T$<$755\,s).  The
patch disappears when the \ud's brightness decreases at around 
T$=$2202\,s.  The vector magnetic field of UD\#C gets more 
inclined, especially during the migration.

\subsubsection{Typical Grain-Origin \uds}\label{sec:grain-origin}

Grain-origin \uds\ are characterized by inward migration to the umbra
center and high intensity contrasts (Table\,\ref{tbl:table1}).  The
perturbations associated with these \uds\ are more clearly visible
than those of central and peripheral \uds.  Our analysis demonstrates
that more than 70\% of the grain-origin \uds\ harbor upflows, while 40--50\%
show weaker and more inclined fields. The magnetic field perturbations
occur preferentially near the penumbra, becoming less prominent as the
UD moves into the umbra.
Sometimes one observes patches of both reduced and increased field
strengths at the position of grain-origin UDs.

In this section we describe the evolution of two typical grain-origin
\uds\/: one from the disk center side of the spot and the other from 
the limb side.

Figures~\ref{fig:PG_094_sequence} and \ref{fig:PG_094_plot} show the
temporal evolution of the disk-center side UD (UD\#D).  It evolves
from a filamentary structure into a circular shape and follows an
unusual trajectory: generally, grain-origin \uds\ move along the
extension line of penumbral grains, but the trajectory of UD\#D is
almost perpendicular to it.  The continuum intensity decreases a bit
during 188\,s$<$T$<$314\,s, and then increases again.  Upflows are
observed in the first half of the UD evolution, weakening until they
almost reach the background level at $T=692$~s. The strongest flow
(\vdeep $< -0.6$~\kms) occurs at the edge of the penumbral grain early
in the UD's evolution, around $T= 62$\,s, and appears to be the
continuation of the typical upflow of penumbral grains.

UD\#D is located at the boundary of weak and strong field regions.
When it moves along the $-y$ direction, the boundary also evolves as
if the leading edge of the \ud\ was always blocked by strong field
walls (0\,s$<$T$<$817\,s). After that, the UD collides with the strong
field region and then disappears together with the strong field walls. 
We could not detect any systematic perturbation of the field
inclination caused by this UD.

Figures~\ref{fig:PG_051_sequence} and \ref{fig:PG_051_plot} show the
temporal evolution of UD\#E, a limb-side UD.  This structure is
detached from a penumbral grain and moves along the extension line of
the grain with an apparent speed of 0.7\,\kms. In the case of
limb-side UDs, velocity perturbations are hard to detect.  UD\#E does
show upflows in \vhigh\ and \vdeep, but they are much weaker than
those observed in center-side UDs.  We attribute this to a
line-of-sight effect working against the field-aligned flows in the
inclined magnetic field. During the first part of its evolution, the
UD is located in between two patches of weaker and stronger fields.
Like in the case of UD\#D, the migration seems to be impeded by strong
field walls. Toward the end of the sequence (at $T=692$~s), the UD
only shows stronger fields than the surroundings. It does not seem to
perturb the inclination of the umbral field.

\subsubsection{\uds\ with Downflow Patches}\label{sec:verydeep}

\citet{2010ApJ...713.1282O} found evidence of downflow patches 
associated with bright \uds\ in a pore.  A bisector analysis of the
\ion{Fe}{1} 6301.5 line showed that these downflows are strongest
in deep atmospheric layers.  We tried to perform a similar analysis,
but this proved difficult for two reasons:
\begin{enumerate}
\item In the darkest parts of our large spot, the \ion{Fe}{1} 6301.5\,\AA\ 
bisectors at high intensity levels ($> 70$\%) appear to show
systematic blueshifts.
\item Near the continuum, the quality of the bisector maps is very
dependent on the seeing conditions.   
\end{enumerate}
Therefore in this section we show examples of downflows extracted from
the best scan in our data set, focussing on relatively bright areas
around the center of the umbra.

\begin{figure}[t]
\centerline{\includegraphics[width=90mm, bb=0 0 425 510]{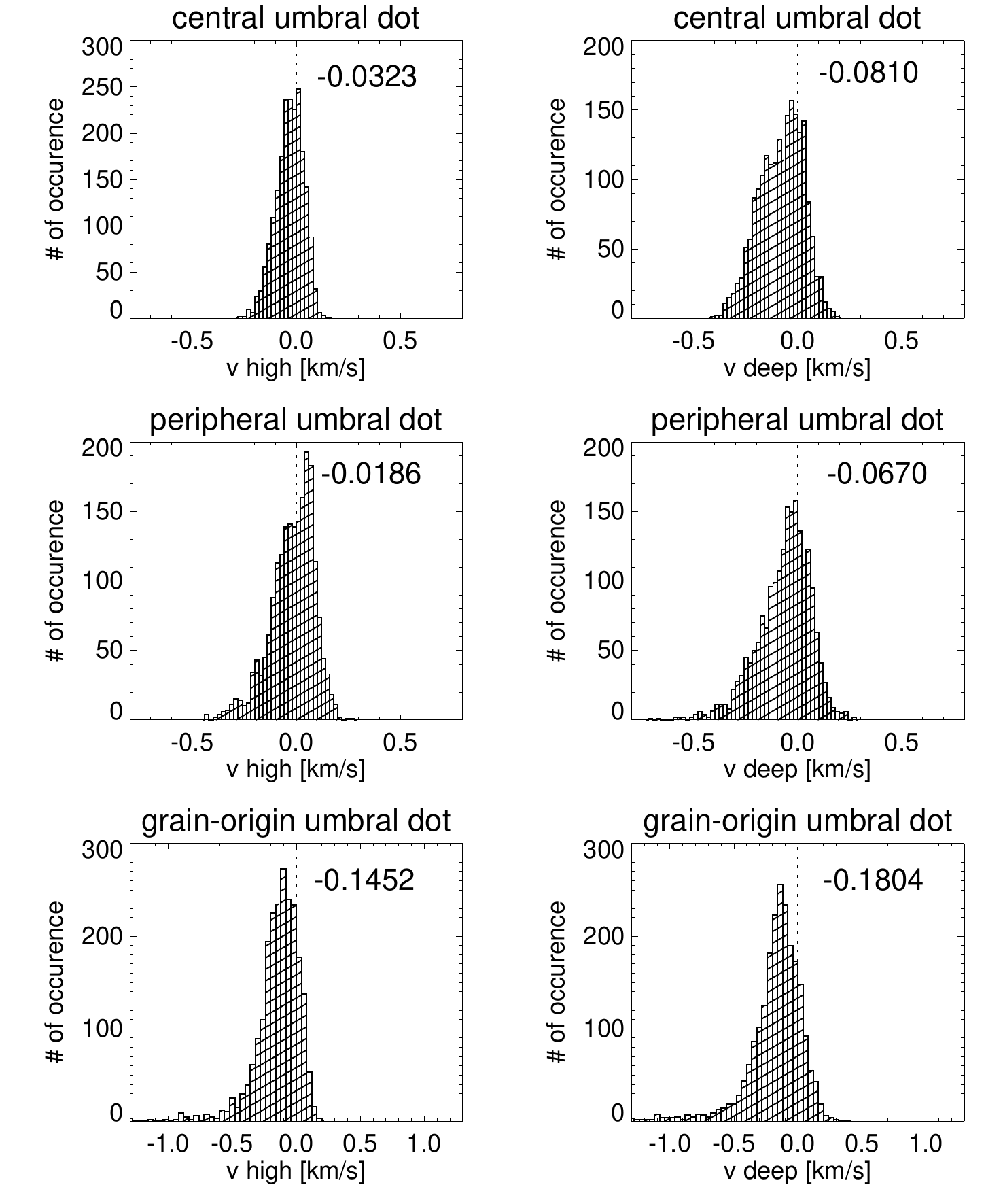}}
   \caption{Histograms of average \vhigh\ and \vdeep\ in central UDs
   (top), peripheral UDs (middle), and grain-origin UDs (bottom). The
   center-of-gravity value of each histogram is given in the upper
   right corner.  }
\label{fig:histogram_bisector}
\end{figure}

Figure~\ref{fig:verydeep} shows a region of about $4\arcsec \times
2\arcsec$ centered at $(x,y)=(33\arcsec, 38.5\arcsec)$.  The two UDs
marked in the upper left panel (UD\#F and UD\#G) exhibit no clear
velocity signals in \vhigh\ and \vdeep, but prominent upflows in the
bisector map at the 90\% intensity level (lower left panel).  Next to
those upflows, localized downflow patches can be seen in the vicinity
of the \uds.  The blueshift at the center of UD\#F amounts to
$-1.3$\,\kms, while the downflow patches have speeds of 0.16\,\kms\
($+y$ side) and 0.75\,\kms\ ($-y$ side).  Similarly, the blueshift 
at the center of UD\#G is $-$1.5\,\kms, and the downflows to the top
right attain 0.7\,\kms. 
The downflow patches have an approximate size of 0\farcs2.  In these
\uds\ we do not see the central dark lanes predicted by
\citet{2006ApJ...641L..73S}.

UD\#F and UD\#G show slightly weaker and more inclined fields than
their surroundings.  When we inspect the continuum movie, these two
UDs are both in the peak phase of their brightness.

\subsection{Statistical Properties}

In this section we describe the statistical properties of 339 \uds\
(98 central, 112 peripheral, and 129 grain-origin \uds). Histograms of
lifetimes, average proper motions, brightness ratios, and diameters
have already been presented in Figure~\ref{fig:histogram_4parameter}.

\subsubsection{Bisector Velocities}

Figure~\ref{fig:histogram_bisector} displays histograms of the mean
\vhigh\ and \vdeep\ in UDs (averaged over an area of $7 \times 7$
pixels centered at the UD's position), including all temporal steps
within their lifetimes (i.e., 6279 samples).  The center-of-gravity
value is given in each panel.  If UDs had no systematic flows, the
histograms would show a symmetric distribution about 0\,\kms.  This is
the case for central \uds, with only a slight inclination to negative
velocities (upflows).  On the other hand, a tail extending to strong
upflows is seen in the histograms for peripheral and grain-origin
\uds.  The asymmetric distribution is more prominent in the case of
grain-origin \uds. Statistically, the upflows are stronger in \vdeep\
compared to \vhigh, indicating a deceleration with height.

\begin{figure}[t]
\centerline{\includegraphics[width=85mm, bb=0 0 340 255]{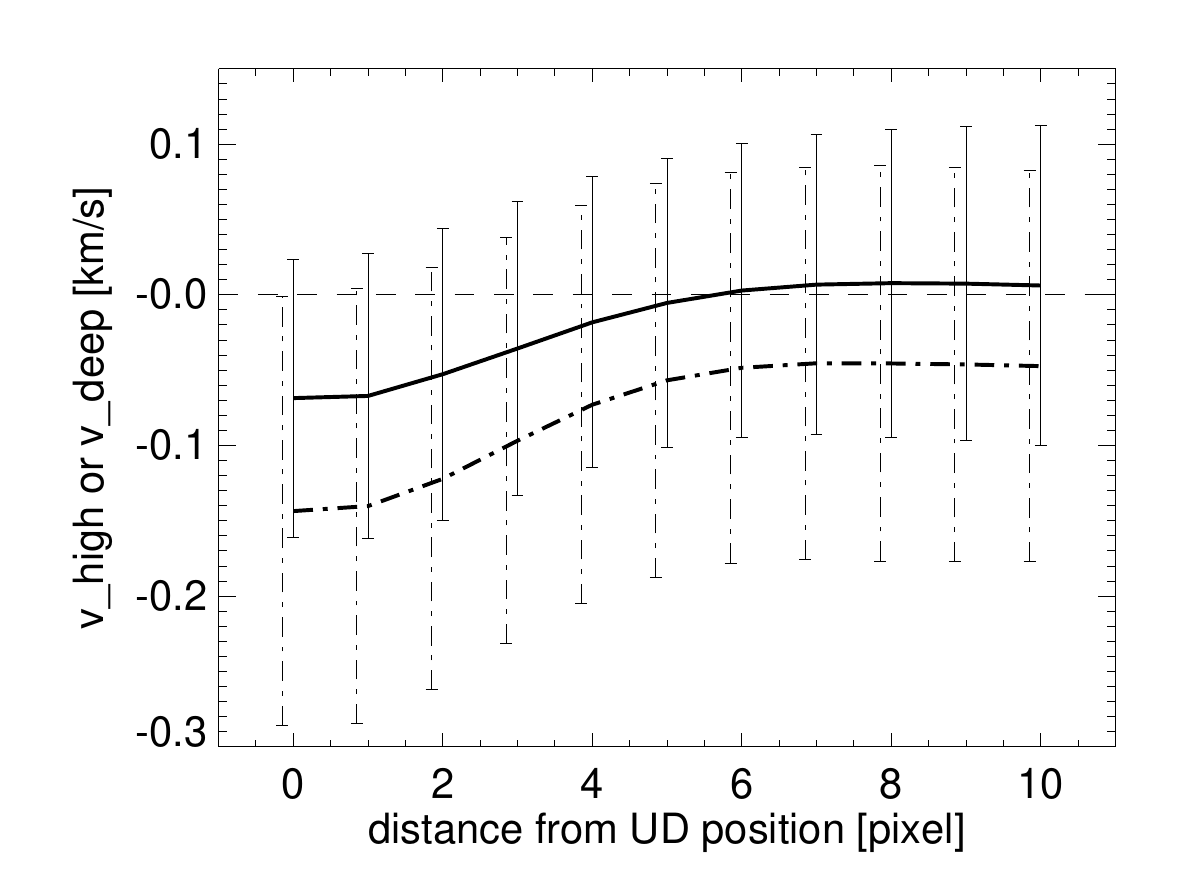}}
   \caption{Average value of \vhigh\ (solid) and \vdeep\ (dash-dotted)
   as a function of radial distance from the UD center.  Negative
   velocities mean upflows. The calculation is performed in the umbra,
   where the continuum intensity is darker than 0.4$ I_{\rm
   qs}$.  One pixel corresponds to 0\farcs06.  The error bars show the
   1$\sigma$ fluctuation within 1 pixel bins.  \vspace{0.2cm}}
   \label{fig:bisector_distance}
\end{figure}

Figure~\ref{fig:bisector_distance} shows the average variation of the
velocity as a function of radial distance from the center of the UD.
Upflows decreasing outward are found in the region close to the
\uds\ ($<$5~pixels), although the 1$\sigma$ fluctuation is large.
The strongest upflows occur at the peak brightness position, with
average velocities of \vhigh$= -0.07$\,\kms\ and \vdeep$=-0.14$\,\kms.
Within a distance of 10~pixels (0\farcs6) from the \ud\ center, we
always find upward velocities on average, but no downflows.

\subsubsection{Magnetic Parameters}

The local perturbations of field strength ($\Delta B$) and field
inclination ($\Delta i$) are obtained by subtracting a smoothed
version of the maps from the original maps themselves. The smoothing
is done with a boxcar of width $20 \times 20$~pixels ($870 \times 870$
\,km$^2$), which is significantly larger than the typical UD size 
(see Figure~\ref{fig:histogram_4parameter}).

\begin{figure}[t]
\centerline{\includegraphics[width=90mm, bb=0 0 425 510]{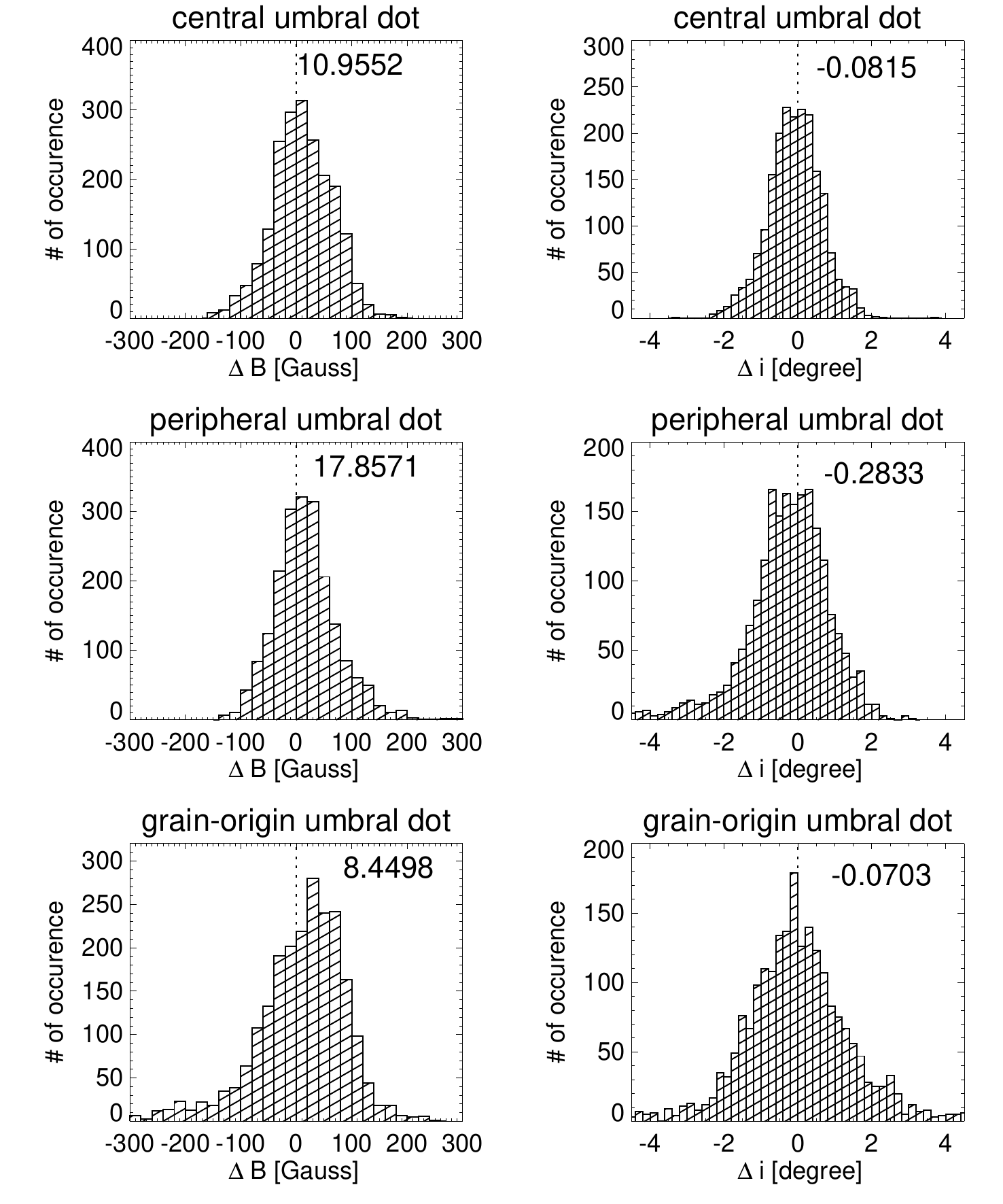}}
   \caption{Histograms of $\Delta B$ and $\Delta i$ for central (top),
   peripheral (middle), and grain-origin (bottom) UDs. These values
   are computed over an area of $7 \times 7$ pixels, centered at the
   position of the UD. Negative $\Delta i$ means more vertical field
   lines, while positive $\Delta i$ means more inclined fields.  Shown
   in the upper right corner is the center of gravity of each
   histogram. 
   } \label{fig:histogram_magnetic_difference}
\end{figure}

Figure~\ref{fig:histogram_magnetic_difference} shows the histograms of
$\Delta B$ and $\Delta i$ averaged over an area of $7 \times 7$ pixels
centered at the UD's position.  To our surprise, the histogram of
$\Delta B$ indicates that the \ud\ magnetic field is a bit stronger
than the surroundings, which seems to contradict the field-free model
of \citet{2006ApJ...641L..73S}.  The histograms of $\Delta i$ are
almost symmetric, indicating no preference for more vertical or more
inclined fields at the position of the UDs.

\subsubsection{Temporal Evolution}

\begin{figure}[t]
\centerline{\includegraphics[width=90mm, bb=0 0 340 566]{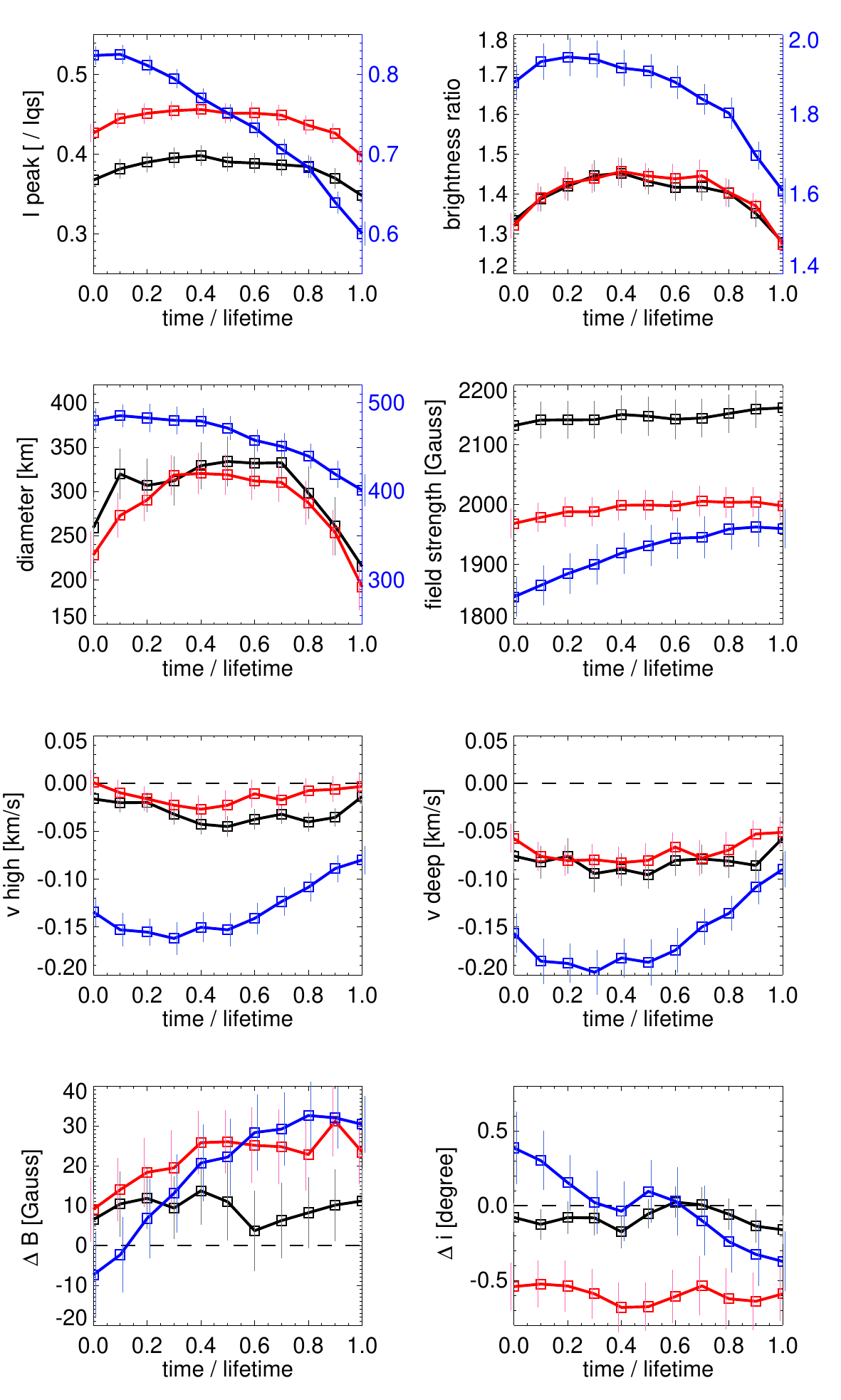}}
   \caption{Temporal evolution of peak brightness ($I_{\rm peak}$),
   brightness ratio, diameter, field strength, \vhigh, \vdeep, $\Delta
   B$, and $\Delta i$ for central (black), peripheral (red), and
   grain-origin (blue) \uds.  The axis for grain-origin \uds\ is
   separately shown to the right in the panels of $I_{\rm peak}$,
   brightness ratio, and diameter. The standard errors of the mean
   (SEM) are indicated by the vertical lines.  \vspace{0.2cm}}
   \label{fig:typical_lightcurve}
\end{figure}

We have studied the temporal evolution of \uds\ using the
ones that were observed from birth to death and had lifetimes longer
than 620\,s. In total, 36 central, 50 peripheral, and 66 grain-origin
\uds\ were chosen for this analysis. By normalizing the lifetimes 
to unity it is possible to average all the evolutionary curves. The
results are shown in Figure~\ref{fig:typical_lightcurve} for eight
parameters: peak brightness ($I_{\rm peak}$), brightness ratio,
diameter, field strength, \vhigh, \vdeep, $\Delta B$, and $\Delta i$. 
The parameters have been averaged over an area of $7 \times 7$ pixels
centered at the UD's position. 

\paragraph{Evolution of Central and Peripheral \uds}

For central and peripheral \uds, the peak brightness, the brightness
ratio, and the diameter show a symmetric increase and decrease over
time. The brighter $I_{\rm peak}$ and weaker field strengths observed
in peripheral \uds\ are a natural consequence of their location in the
more external parts of the umbra.  Both central and peripheral UDs
harbor upflows (negative \vhigh\ and \vdeep) that grow with time,
reach a maximum when the UDs are mature, and then decrease more or
less symmetrically. The brightness follows a similar pattern.  Thus,
there is a positive correlation between brightness and upflows, which
is a sign of convection.

The field strength and $\Delta B$ are nearly constant, although
$\Delta B$ shows a tendency to increase during the evolution of
peripheral UDs.  $\Delta i$ is around zero for central
\uds\ and negative for peripheral UDs (indicating more vertical
fields), with little variations over time.

\paragraph{Evolution of Grain-Origin \uds}

For grain-origin \uds, all parameters other than bisector velocities
show patterns of monotonic increases or decreases.  These patterns are
caused by the smooth transition from penumbral grains to circular
\uds\ in the central umbra.  The bisector velocities peak shortly 
after the appearance of the UDs, when the tips of penumbral grains are
completely detached.  Although with large scatter, locally weaker and
more inclined fields are found in the first half of the UD's lifetime,
while opposite properties (locally enhanced and more vertical fields)
appear in the latter half.

\subsubsection{Scatter Relations}\label{sec:scatter}

\begin{figure*}[bhtp]
\centerline{\includegraphics[width=0.87\textwidth, bb=0 0 708 566]{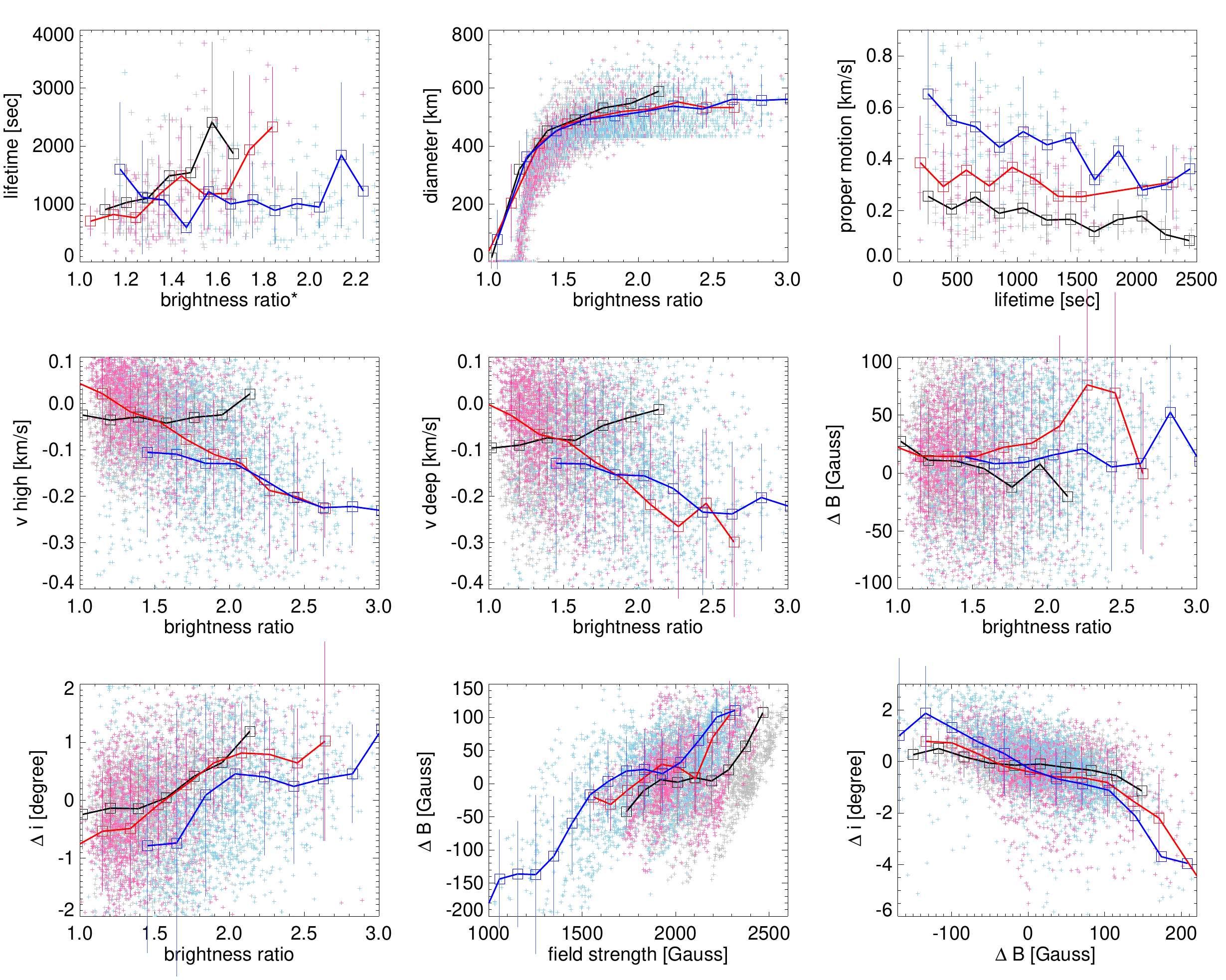}}
\caption{Scatter plots of brightness ratio, lifetime, proper motion, 
diameter, field strength, \vhigh, \vdeep, $\Delta B$, and $\Delta i$,
for central (black), peripheral (red), and grain-origin (blue) UDs.
The squares and the thick lines indicate the average values in the
corresponding bins.  The 1-$\sigma$ fluctuations within the bins are
shown by vertical lines. The brightness ratio used in the top left
panel is an average over the UD's lifetime. }
\label{fig:scatter_9parameter}
\end{figure*}

Scatter plots of various parameters are displayed in
Figure~\ref{fig:scatter_9parameter}, separately for central,
peripheral, and grain-origin UDs (black, red, and blue, respectively).
In general, all types of UDs exhibit similar relations between
parameters, although there are some differences in bisector
velocities. Figure~\ref{fig:scatter_9parameter} shows that:

\begin{enumerate}
\item Brighter UDs tend to have longer lifetimes. This is because 
      most of them are recurrent. Other than that, the lifetime is
      almost constant regardless of the type of UD.

\item The diameter increases linearly for  brightness ratios up to 1.4 
      due to the way it is calculated. For larger ratios the diameter
      saturates to a constant value of about 500~km, demonstrating that
      UDs have a typical size.

\item The proper motion speed decreases linearly with lifetime: 
      short-lived UDs move faster.
  
\item For peripheral and grain-origin UDs, the stronger upflows are found 
      in the brighter structures.  Central UDs show the opposite
      tendency.  
 
\item There is no correlation between $\Delta B$ and brightness ratio.
 
\item Brighter UDs exhibit more inclined fields than dark \uds, although 
      the inclination difference is not larger than 2$^{\circ}$ on average.

\item There is a clear correlation between $\Delta B$ and field strength. 
      For UDs with strong fields ($>$2000~G), $\Delta B$ is positive
      and reaches up to $\approx$100~G.

\item The most inclined fields are associated with negative $\Delta B$ 
      perturbations. UDs with more vertical fields tend to show positive 
      $\Delta B$ values.
\end{enumerate}

Bisector velocities, $\Delta B$, and $\Delta i$ are the most important
parameters to decide on the convective nature of \uds.  The field-free
convection model of \uds\ suggests upflows, weaker, and more inclined
magnetic fields.  These conditions are indeed true for relatively weak
UDs ($<2000$~G). However, strongly magnetized UDs ($>2000$~G) usually
show positive $\Delta B$ perturbations and more vertical field
lines. We conjecture that the stronger upflows observed
in the darker central \uds\ may be the result of contamination by
molecular lines.

\section{Discussion and Conclusion}\label{sec:discussion}

In this paper we have performed a detailed analysis of \uds\ in a
mature sunspot using data from the CRISP spectropolarimeter.  The
excellent spatial resolution, temporal cadence, and polarimetric
sensitivity of the measurements are ideal for \ud\ studies.  The
perturbations caused by \uds\ are usually very small, and thus an
statistical approach has to be followed to reveal their common
properties.  Our work addresses for the first time the temporal
evolution of velocity and magnetic fields in and around \uds, using a
statistically significant sample of \uds.

\paragraph{Convection in the umbra}

\Uds\ are considered to be the manifestation of convection 
in the presence of the strong umbral field, while more vigorous
convection occurs in the quiet Sun in the form of granules.  The
morphological differences between granules and \uds\ can be seen in
Figure~\ref{fig:convection_morphology}. Driven by overshooting
cellular convection, granules are characterized by sharp edges and
irregular polygonal shapes \citep{1990ARA&A..28..263S}.
\Uds, on the other hand, show Gaussian brightness profiles. 
Linear theory reveals that the preferred horizontal scale of
convection decreases with increasing field strength
\citep{1990MNRAS.245..434W}, which explains why \uds\ are smaller
than granules.  The convective origin of \uds\ seems well established,
as many papers including ours found a good correlation between upflows
and brightness.  However, in the umbra there are other diffuse areas
with enhanced brightness whose origin is still unknown. Our
speculation that the diffuse bright areas of the umbra are also caused
by convection could not be unambiguously confirmed on the basis of a
unique relation between brightness and velocities (Figure~\ref{fig:bisector_scatter}).
  
\paragraph{Photometric properties} 

\citet{2009ApJ...702.1048W} found constant lifetimes regardless of the \ud\  
type and the structure of magnetic field at the position of the UD.
This is in agreement with our results (Section\,\ref{sec:scatter}).
However other studies report longer lifetimes for brighter \uds\
\citep{2002A&A...388.1048T, 2010A&A...510A..12B}, and we speculate
this is partly due to the fact that bright \uds\ tend to reoccur
at the same position. 
 
A correlation between shorter lifetime and faster proper motion is
reported for the first time in this paper
(Figure~\ref{fig:scatter_9parameter}). If the energy dissipation rate
is proportional to the UD speed, the lifetime can be expected to be
reduced for fast-moving UDs, as observed. We also find that the
travelled distance depends linearly on lifetime, i.e., long-lived
\uds\ travel longer distances even though they move at lower speeds.
The same conclusion can be obtained from a similar analysis of the
data set of \citet{2009ApJ...702.1048W}.

We found \ud\ diameters consistent with the values reported in the
literature, i.e., about 400\,km on average.  The Gaussian shape of the
histogram (Figure~\ref{fig:histogram_4parameter}) and the lack of
dependence of the diameter on brightness ratio
(Figure~\ref{fig:scatter_9parameter}) suggest that \uds\ indeed have a
``typical'' size, regardless of their type.  This common
\ud\ size is probably determined by a universal near-surface
stratification in mature sunspots. However, the scatter plot analysis
performed in Section~\ref{sec:scatter} did not reveal any physical
parameter having a strong correlation with the \ud\ size.

The intensity oscillations in UDs reported by, e.g., 
\citet{1997ApJ...490..458R} have not been studied in this paper
because of the uncertainties that residual seeing fluctuations may
introduce. However it is true that many \uds\ show recurrence, as
observed also by
\citet{2012ApJ...752..109L}. For example, the peripheral UD\#C
displayed in Figure~\ref{fig:PUD_036_plot} reappeared twice within a
time interval of 13~minutes.  This timescale is comparable to the
oscillatory period of the \uds\ shown in
\citet{1997ApJ...490..458R} and \citet{2009PASJ...61..193W}.

\paragraph{Categorization of \uds}

We classified the observed \uds\ in central, peripheral, and
grain-origin UDs according to their place of birth.  Do these
categories represent physically different structures or different
manifestations of the same phenomenon?  Grain-origin
\uds\ have larger brightness ratios, larger sizes, and faster proper
motions than the other UDs.  The temporal evolution of grain-origin
\uds\ is smooth and shows monotonic changes, while central and
peripheral
\uds\ show mound-shaped evolutionary curves
(Figure~\ref{fig:typical_lightcurve}).  Despite this, the scatter
plots presented in Figure~\ref{fig:scatter_9parameter} suggest that
the properties of grain-origin \uds\ lie on the extension
lines of those of central and peripheral \uds.  The differences
between them may arise from stronger convection in grain-origin \uds\
rendered possible by the weaker background
field. \citet{2006A&A...447..343S} speculated that field-free
convection can explain both \uds\ and penumbral grains. The computer
simulations of \citet{2007ApJ...669.1390H} and
\citet{2009ApJ...691..640R} succeeded in reproducing basic properties
of penumbral filaments and \uds\ as weakly-magnetized convective
structures.  Our results are in general agreement with the predictions
of this scenario, although they also indicate that UDs are far from
being completely field-free (at least in the photospheric layers
accessible to the observations).

\paragraph{Substructures}

Localized downflow patches at the periphery of UDs are considered to
be a signature of overturning convection in \uds\
\citep{2006ApJ...641L..73S}.  In Section\,\ref{sec:verydeep} we 
presented some examples of downflow patches without performing a full
statistical analysis.  The patches of Figure~\ref{fig:verydeep} have
sizes of 0\farcs2 and redshifts of up to 0.75\,\kms. Both the size and
the velocity are in good agreement with those reported by
\citet{2010ApJ...713.1282O} in a pore.  The fact that the downflow
patches are observable only at very high bisector levels (i.e., deep
photospheric layers) is also consistent with the results of those
authors.

However, many \uds\ do not show downflow patches in our data. This
lack of detection could be due to:
\begin{enumerate}
\item Insufficient spatial resolution. 
\item The existence of downflows only in deep layers that cannot be 
probed by the \ion{Fe}{1} 6301 and 6302 \AA\/ lines.
\item The transient nature of the downflows, which could appear only in 
       a particular phase of the UD's evolution.
\item The possibility that the convective energy escapes to the upper 
      layers instead of returning to deep layers \citep[see the narrow
      jet-like upflows above the cusp in][]{2006ApJ...641L..73S}.
\end{enumerate}

The two \uds\ featured in Figure~\ref{fig:verydeep} are in the peak
brightness phase when they show downflows.  Possibly the speed of
the downflows reaches a maximum when the brightness is also maximum.  

\paragraph{Velocities in \uds} 

Upflows and brightness follow similar evolutionary patterns in UDs
(Figure~\ref{fig:typical_lightcurve}).  This correlation supports the
convective nature of \uds\ \citep{2009ApJ...694.1080S,
2010SoPh..266....5W}.  The scatter plots of velocity vs brightness
shown in Figure~\ref{fig:scatter_9parameter} also point to a
convective origin of UDs. For central UDs, however, we observe
stronger blueshifts in darker structures, although the tendency is not
very pronounced. We suspect this is an artifact caused by systematic
blueshifts in very cold umbral areas.

The velocities observed within \uds\ are likely to represent
field-aligned flows, because upflows are readily found on the
disk-center side where the field is closer to the line-of-sight
direction, compared to the limb side
(Section\,\ref{sec:grain-origin}).  The same effect can be observed in
the Evershed flow (Figure~\ref{fig:data_reduction}), which is also a
field-aligned flow.

\paragraph{Magnetic field in and around \uds}

The field-free convection model of \uds\ \citep{2006ApJ...641L..73S}
predicts weaker and more inclined fields in \uds. Our scatter analysis
(Figure~\ref{fig:scatter_9parameter}) confirms these properties, but
only for \uds\ with fields below 2000\,G.  In strongly magnetized 
\uds\ ($>2000$~G), the magnetic field is enhanced and more vertical
compared to the surroundings.  For grain-origin UDs, the physical
conditions also depend on the phase of evolution: in the first half of
their lifetime, weaker and more inclined fields appear, while stronger
and more vertical fields are observed in the latter half as the UDs
intrude into the umbra.  To the best of our knowledge, the enhanced
and more vertical fields of strongly magnetized \uds\ cannot be
explained by currently available \ud\ models.

We observe strong field regions at the migration front of grain-origin
UDs for the first time (see the evolution of UD\#D and UD\#E in
Section~\ref{sec:grain-origin}).  These strong field regions seem to
impede the migration of the UDs. The situation is reminiscent of that
modeled by
\citet{1998A&A...337..897S}, where a weakly magnetized penumbral
flux tube pushes and compresses the pre-existing vertical field at the
leading edge.  However, also a weakly magnetized convective structure
would produce a compression of the adjacent magnetic field, which has
to wrap around the field-free gas. The MHD simulations performed by
\citet{2007ApJ...669.1390H} predict the existence of enhanced field
regions surrounding grain-origin \ud\ only in layers deeper than the
continuum forming region (see Figure 3 in that paper), but on both the
leading and the tail sides.  Our observation did not find field
enhancements on the tail side.

\paragraph{Final remarks} 

This work extends our knowledge of the temporal evolution of
velocities and magnetic fields in UDs.  We found some new and
unanswered results that may provide constraints to future 
modeling efforts.  A pioneering comparison of observational and
computer-simulated \uds\ has been performed by
\citet{2010A&A...510A..12B}, and this kind of studies should be
extended.

At the same time, more spectropolarimetric observations of UDs at high
cadence should be performed.  The temporal resolution of our data,
63\,s, seems appropriate to track the evolution of \uds, but is
insufficient for resolving the evolution of UD substructures. \Uds\
will remain one of the most challenging targets for solar observations
in the coming years.

\acknowledgments 

This work was started while HW was a Visiting Scientist at Instituto
de Astrof\'{\i}sica de Andaluc\'{\i}a.  The Swedish 1-m Solar
Telescope is operated on the island of La Palma by the Institute for
Solar Physics of the Royal Swedish Academy of Sciences in the Spanish
Observatorio del Roque de los Muchachos of the Instituto de
Astrof{\'\i}sica de Canarias. Our work was supported by the
Grant-in-Aid for JSPS Fellows, and by the Grant-in-Aid for the Global
COE Program ``The Next Generation of Physics, Spun from Universality
and Emergence'' from the Ministry of Education, Culture, Sports,
Science and Technology (MEXT) of Japan. We gratefully acknowledge
financial support from the Spanish Ministerio de Ciencia e
Innovaci\'on through projects AYA2009-14105-C06-06 and
PCI2006-A7-0624, and from Junta de Andaluc\'{\i}a through project
P07-TEP-02687, including a percentage from European FEDER funds. This
research has made extensive use of NASA's Astrophysical Data System.

\bibliographystyle{apj}
\bibliography{hw}

\newpage

\begin{figure*}[bhtp]
\centerline{\includegraphics[width=150mm, bb=0 0 708 708]{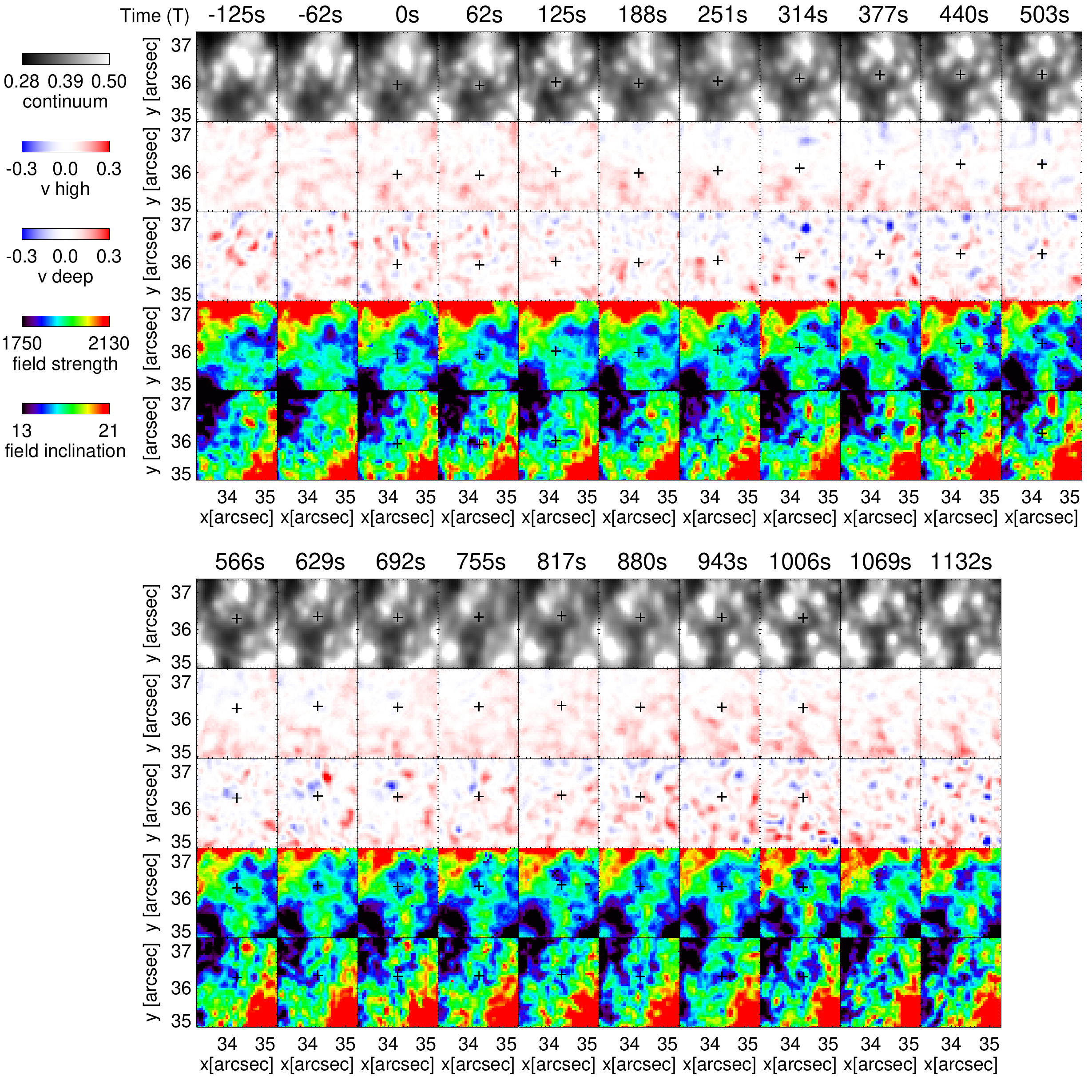}}
   \caption{Temporal evolution of a central \ud\ (\#A) from 2 frames
   before appearance until 2 frames after disappearance.  From top to
   bottom, maps of continuum intensity, \vhigh,
   \vdeep, field strength, and field inclination are shown.
   $T =0$ s corresponds to the appearance of the \ud.  The black plus symbols
   indicate the UD position.  The $(x,y)$ coordinates follow
   the coordinate system used in Figure~\ref{fig:fig1}.}
   \label{fig:centralUD_34_sequence}
\end{figure*}

\begin{figure*}[bhtp]
\centerline{\includegraphics[width=150mm, bb=0 0 595 765]{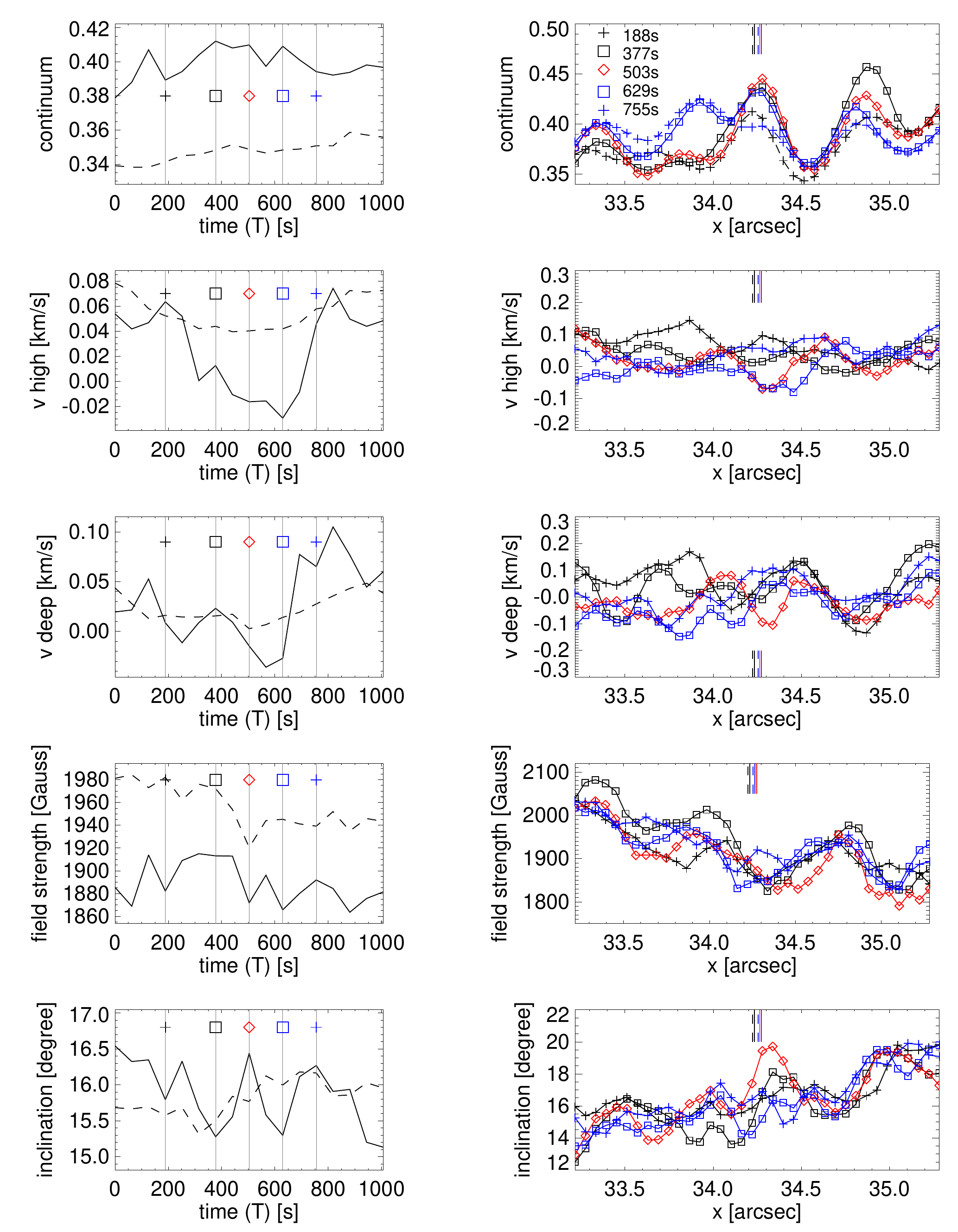}}
   \caption{{\it Left}: Temporal variation of continuum intensity,
   \vhigh, \vdeep, field strength, and field
   inclination for central \ud\#A (solid line, average within
   \ud's position $\pm$ 2 pixel area).  The dashed lines correspond to
   the dark background.  Time starts from the appearance of the \ud.
   {\it Right}: Cuts along the $x$-direction of continuum intensity,
   \vhigh, \vdeep, field strength, and field
   inclination. The five different lines mark the following times:
   188\,s (black plus, dashed line), 377\,s (black square), 503\,s
   (red diamond), 629\,s (blue square), and 755\,s (blue plus, dashed
   line).  The short vertical lines at the upper or lower x-axes 
   indicate the \ud's $x$-axis positions in those five instants.}
   \label{fig:centralUD_34_plot}
\end{figure*}

\begin{figure*}[bhtp]
\centerline{\includegraphics[width=150mm, bb=0 0 708 1105]{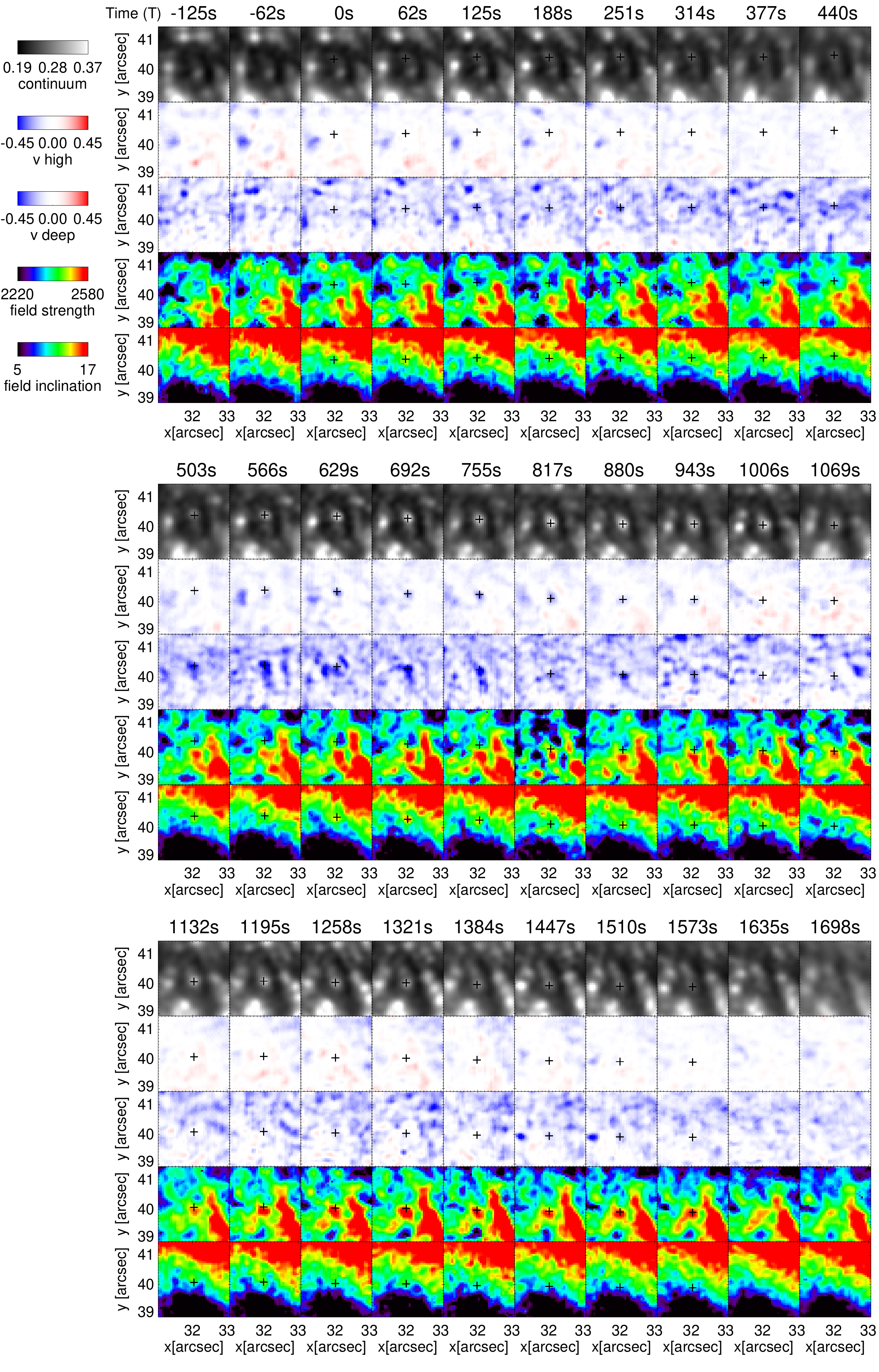}}
   \caption{Same as Figure~\ref{fig:centralUD_34_sequence}, for
   central UD\#B.}  \label{fig:centralUD_42_sequence}
\end{figure*}

\begin{figure*}[bhtp]
\centerline{\includegraphics[width=150mm, bb=0 0 595 765]{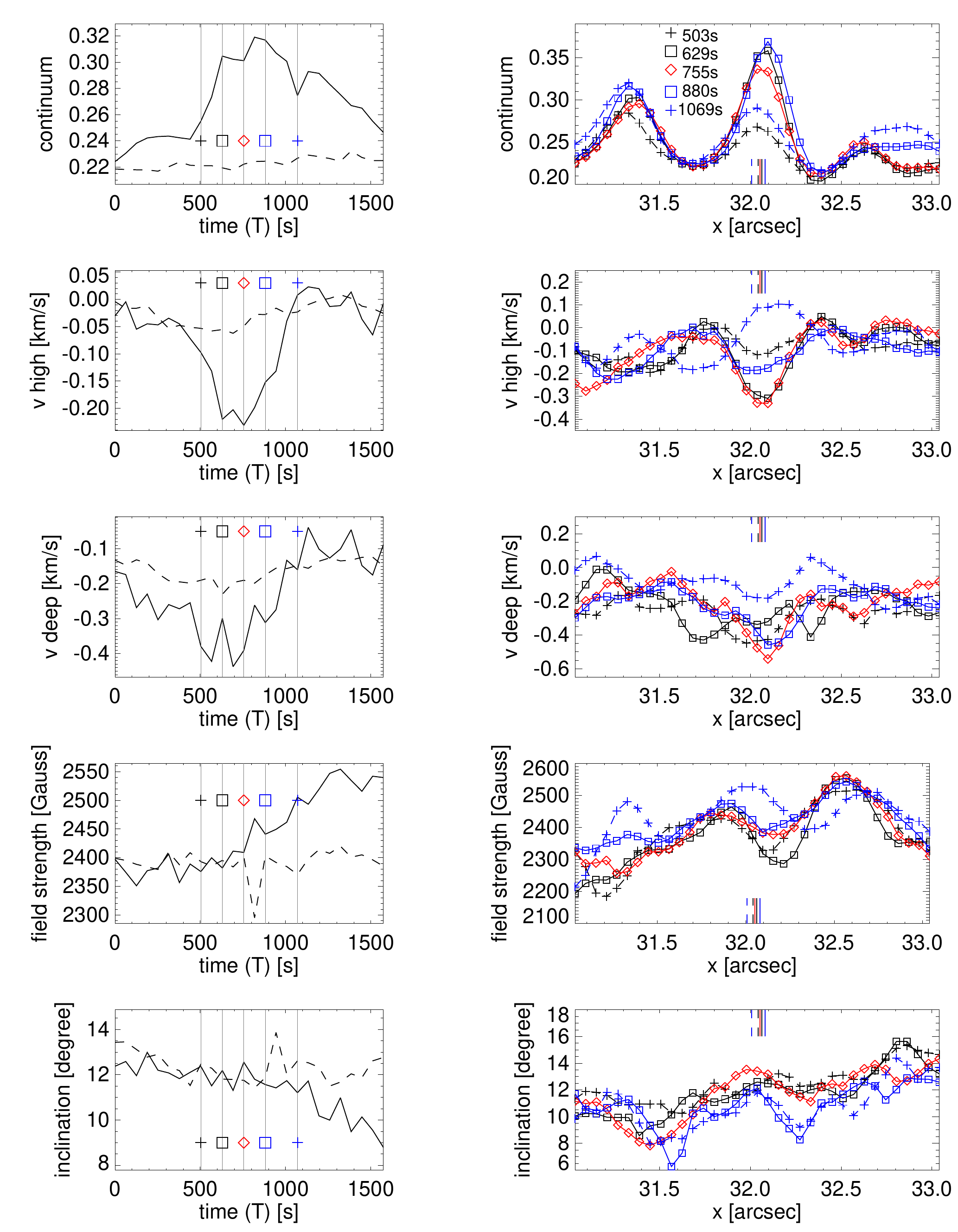}}
   \caption{Same as Figure~\ref{fig:centralUD_34_plot}, for central
   UD\# B.  The times corresponding to the five different lines in the
   right panels are 503\,s (black plus, dashed line), 629\,s (black
   square), 755\,s (red diamond), 880\,s (blue square), and 1069\,s
   (blue plus, dashed line). }  \label{fig:centralUD_42_plot}
\end{figure*}

\begin{figure*}[bhtp]
\centerline{\includegraphics[width=150mm, bb=0 0 708 1038]{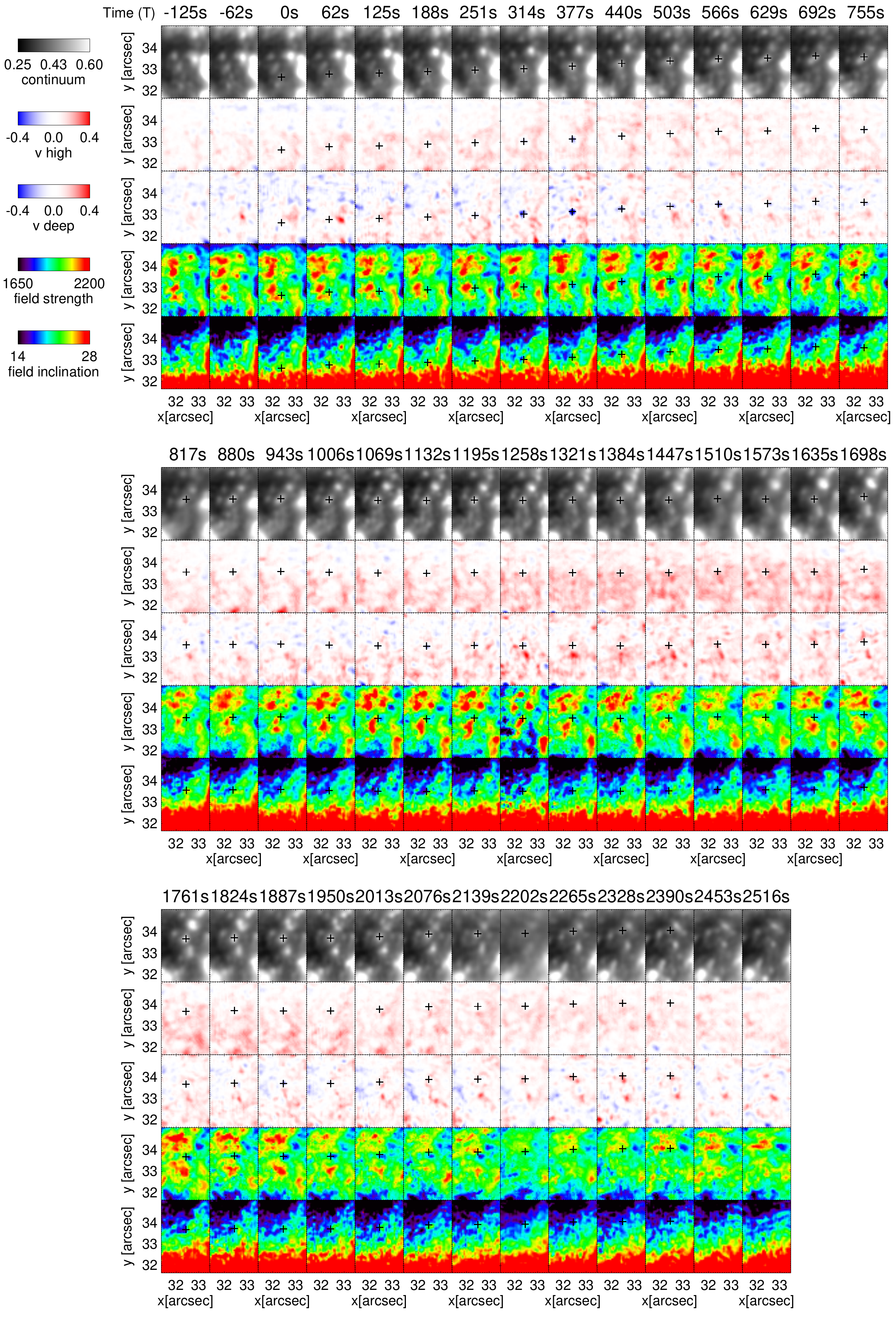}}
   \caption{Same as Figure~\ref{fig:centralUD_34_sequence}, for
   peripheral UD\#C} \label{fig:PUD_036_sequence}
\end{figure*}

\begin{figure*}[bhtp]
\centerline{\includegraphics[width=150mm, bb=0 0 595 765]{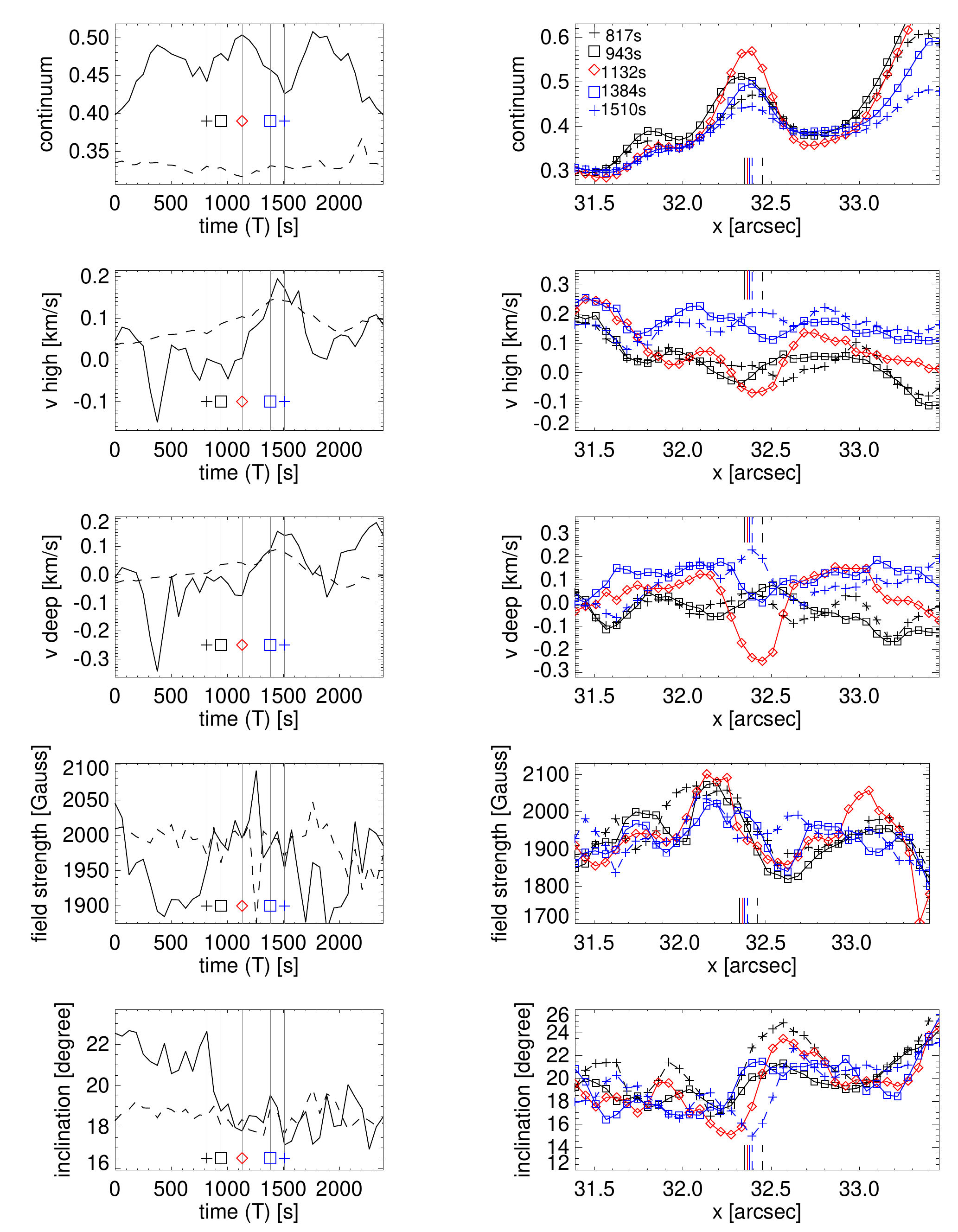}}
   \caption{Same as Figure~\ref{fig:centralUD_34_plot}, for peripheral
   \ud\#C. The times corresponding to the five different lines in the
   right panels are 817\,s (black plus, dashed line), 943\,s (black
   square), 1132\,s (red diamond), 1384\,s (blue square), and 1510\,s
   (blue plus, dashed line). }  \label{fig:PUD_036_plot}
\end{figure*}

\begin{figure*}[bhtp]
\centerline{\includegraphics[width=130mm, bb=0 0 651 1247]{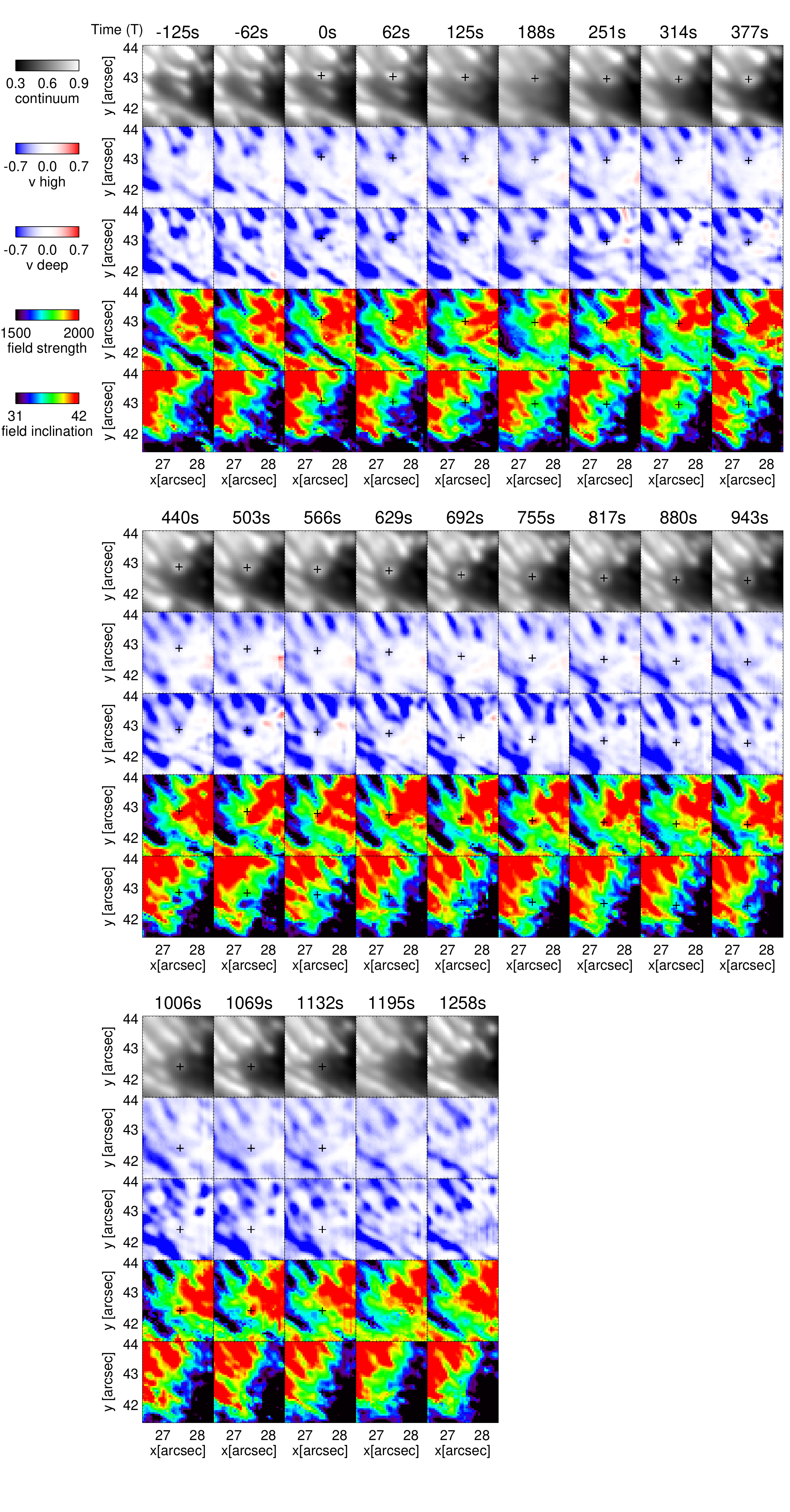}}
   \caption{Same as Figure~\ref{fig:centralUD_34_sequence}, for
   grain-origin UD\#D}. 
	\label{fig:PG_094_sequence}
\end{figure*}

\begin{figure*}[bhtp]
\centerline{\includegraphics[width=150mm, bb=0 0 595 765]{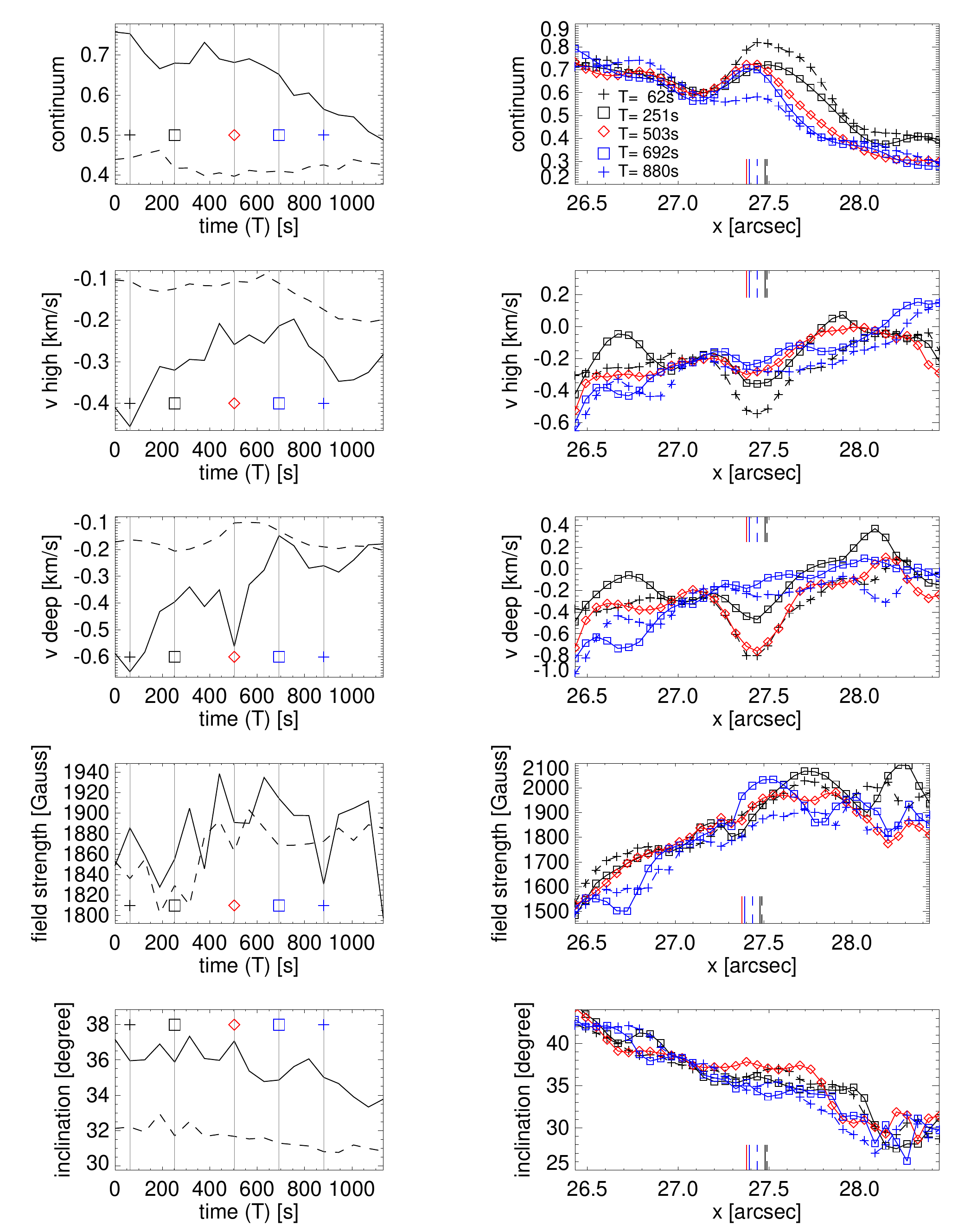}}
   \caption{Same as Figure~\ref{fig:centralUD_34_plot}, for
   grain-origin UD\#D. The times corresponding to the five different
   lines in the right panels are 62\,s (black plus, dashed line),
   251\,s (black square), 503\,s (red diamond), 692\,s (blue square),
   and 880\,s (blue plus, dashed line).}  \label{fig:PG_094_plot}
\end{figure*}

\begin{figure*}[bhtp]
\centerline{\includegraphics[width=150mm, bb=0 0 623 1048]{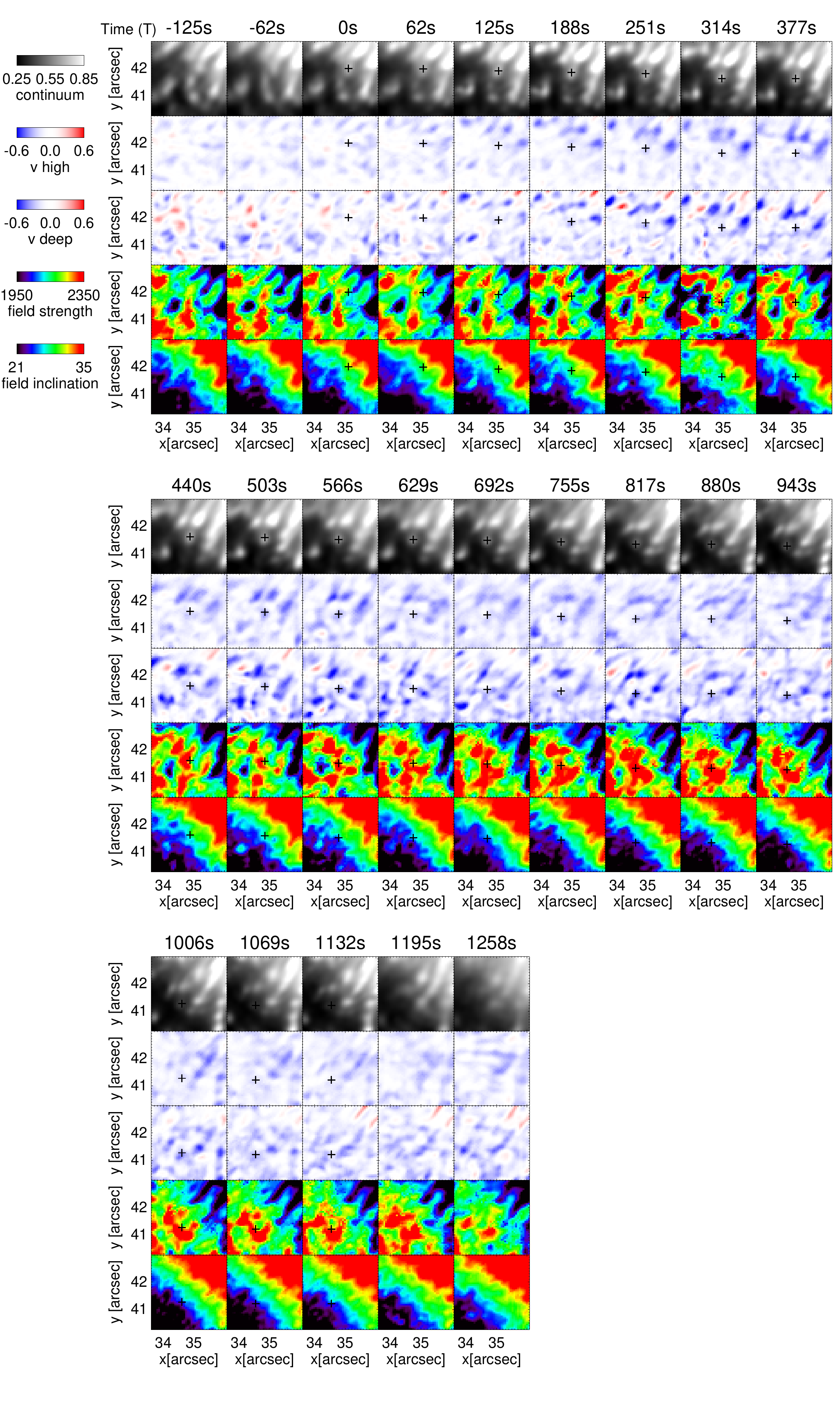}}
   \caption{Same as Figure~\ref{fig:centralUD_34_sequence}, for
   grain-origin UD\#E. 
    }
   \label{fig:PG_051_sequence}
\end{figure*}

\begin{figure*}[bhtp]
\centerline{\includegraphics[width=150mm, bb=0 0 595 765]{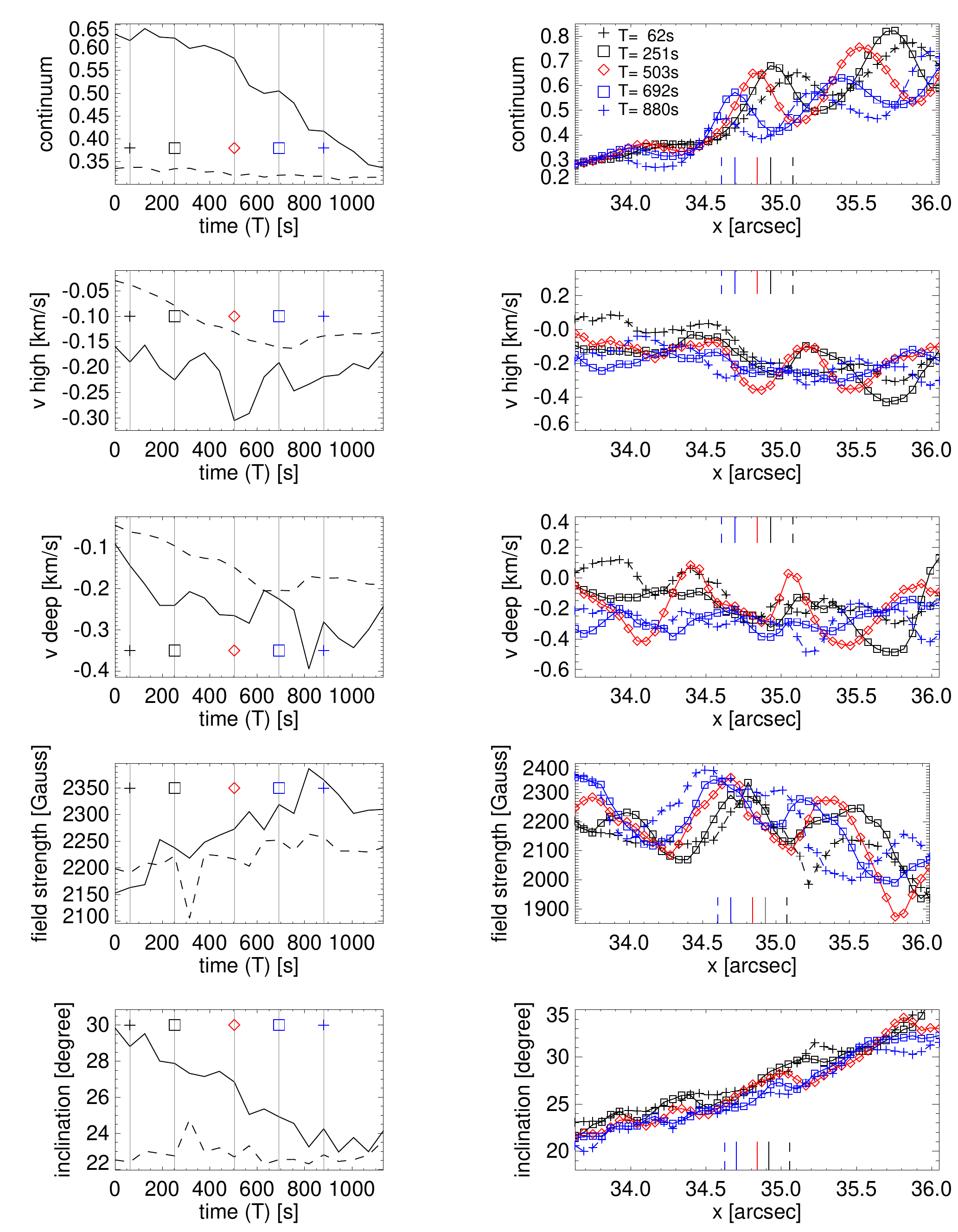}}
   \caption{Same as Figure~\ref{fig:centralUD_34_plot}, for
   grain-origin \ud\#E. The times corresponding to the five different
   lines in the right panels are 62\,s (black plus, dashed line),
   251\,s (black square), 503\,s (red diamond), 692\,s (blue square),
   and 880\,s (blue plus, dashed line). \label{fig:PG_051_plot}}
\end{figure*}
   
\end{document}